\documentclass[twocolumn,showpacs,amsmath,amssymb]{revtex4}
\usepackage{graphicx}
\usepackage{dcolumn}
\usepackage{bm}
\newcommand{\beq}{\begin{equation}}
\newcommand{\eeq}{\end{equation}}
\newcommand{\bea}{\begin{eqnarray}}
\newcommand{\eea}{\end{eqnarray}}
\newcommand{\eps}{\varepsilon}
\newcommand{\bfg}{\boldsymbol}

\begin{document}

\author{E. E. Saperstein}
\affiliation{Kurchatov Institute, 123182 Moscow}
\author{O. Achakovskiy}
\affiliation{Institute for Physics and Power Engineering, 249033
Obninsk, Russia}
\author{S. Kamerdzhiev}
\affiliation{Institute for Physics and Power Engineering, 249033
Obninsk, Russia}
\author{S. Krewald}\affiliation{Institut f\"ur Kernphysik, Forschungszentrum J\"ulich,
D-52425 J\"ulich, Germany}
\author{J. Speth}\affiliation{Institut f\"ur Kernphysik, Forschungszentrum J\"ulich,
D-52425 J\"ulich, Germany}
\author{S. V. Tolokonnikov}
\affiliation{Kurchatov Institute, 123182 Moscow} \affiliation{Moscow
Institute of Physics and Technology, 141700 Dolgoprudny, Russia.}

\title{Phonon coupling effects in magnetic moments  of magic and semi-magic nuclei}

\pacs{21.60.Jz, 21.10.Ky, 21.10.Ft, 21.10.Re}

\begin{abstract} Phonon coupling (PC) corrections to magnetic moments of odd neighbors of
magic and semi-magic nuclei  are analyzed within the self-consistent
Theory of Finite Fermi Systems (TFFS) based on the Energy Density
Functional by Fayans {\it et al}.  The perturbation theory in
$g_L^2$ is used where $g_L$ is the phonon-particle coupling vertex.
A model is developed with separating  non-regular PC contributions,
the rest is supposed to be regular and included into the standard
TFFS parameters. An ansatz is proposed to take into account the
so-called tadpole term which ensures the total angular momentum
conservation with $g_L^2$ accuracy. An approximate method is
suggested to take into account higher order  terms in $g_L^2$. Calculations
are carried out for four odd-proton chains, the odd Tl, Bi, In and
Sb ones. Different PC corrections strongly cancel each other. In the
result, the total PC correction to the magnetic moment in magic
nuclei is, as a rule, negligible. In non-magic nuclei considered it
is noticeable and, with only one exception, negative. On average it
is of the order of $-(0.1 \div 0.5)\;\mu_N$ and improves the
agreement of the theory with the data. Simultaneously we calculated
the gyromagnetic ratios $g_L^{\rm ph}$ of all low-lying phonons in
$^{208}$Pb.  For the $3^-_1$ state it is rather close to the
Bohr--Mottelson  model prediction whereas for other $L$-phonons, two
$5^-$ and six positive parity states, the difference from the
Bohr--Mottelson values is significant.

\end{abstract}

\maketitle

\section{Introduction}

This work is devoted to studying  effects of the interaction of
low-lying collective states (``phonons'') with  single-particle
degrees of freedom. The first microscopic description of phonons in
spherical nuclei was made more than 50 years ago by Spartak Belyaev
in the seminal article \cite{Bel1}.

In the last decade, a considerable success of different approaches
based on the mean field theory in the description of  nuclear bulk
properties was achieved. It concerns the Hartree--Fock (HF) method
with Skyrme force  \cite{HFB-17,EKR},  the HF method
with Gogny force \cite{Gogny}, the Relativistic Mean-Field
approach \cite{RHF} and the so-called Generalized Energy Density
Functional (EDF) method by Fayans et al. \cite{Fay}. The last
approach is essentially close to the self-consistent Theory of
Finite Fermi Systems (TFFS) \cite{KhS} based on TFFS by Migdal
\cite{AB1} supplemented with the Fayans--Khodel self-consistency
relation \cite{Fay-Khod}. A specific feature of the Fayans method is
the consideration of the pairing problem by  solving the Gor'kov
equations in the coordinate space with the method developed by
Belyaev {\it et al.} \cite{Bel2}. Recently, all these approaches were also successfully applied to the description
of phonons in spherical nuclei within the self-consistent
Quasiparticle Random Phase Approximation (QRPA) or similar methods,
see e.g. \cite{Bertsch1,Vor,Bertsch2,RHF-RPA,BE2}.

The next step consists in taking into account the phonon coupling (PC) effects.
These effects were studied in detail in the Quasiparticle-Phonon Model by Soloviev \cite{solov}  and, on the
phenomenological level,  within the so-called Nuclear Field Theory by Bortignon, Broglia
{\it et al.} \cite{NFT}. On a more microscopic level, they
were considered mainly for describing different kinds of Giant Resonances and the
Pygmy-dipole Resonance within the (Q)RPA+PC method \cite{colo1994} and Extended Theory of Finite Fermi
Systems (ETFFS) \cite{revKST}.
  Self-consistent extensions of these methods
  have been proposed recently, for example, (Q)RPA+PC \cite{sarchi} and
  the ETFFS in the Quasiparticle Time Blocking Approximation (QTBA) \cite{tsel2007}
[ETFFS(QTBA)] \cite{avd2007}, see also  the recent review \cite{Pygmy}
devoted to the Pygmy-dipole Resonance. It should be mentioned as well
 successful applications of the self-consistent  ETFFS(QTBA) in \cite{PRL}
and in \cite{AGKK}.   The approach is limited to magic and semi-magic
nuclei where there is a small parameter $g_L^2$, the
 square of the phonon creation amplitude.

Here we concentrate on the self-consistent description of the PC corrections to magnetic moments.
Due to modern Radioactive Ion Beam facilities a lot of new data on magnetic and quadrupole moments
appeared recently including nuclei distant from the beta-decay valley.
 The bulk of the data obtained till 2005 is
collected in a very  comprehensive compilation by Stone
\cite{Stone}. More recent data are presented in original articles,
e.g. in Ref. \cite{Q_Cu},  where new, together with older, data on
magnetic and quadrupole moments of a long chain of copper isotopes
from $N=28$ to $N=46$ are presented and successfully analyzed within
the Many-Particle Shell Model (MPSM) \cite{BABr}. This approach is
very comprehensive as it takes into account all main inter-nuclear
correlations. The necessity to introduce many parameters  of the
effective interaction, the single-particle mean field and the
effective particle charges is some deficiency of the MPSM. In
addition, the domain of the MPSM applications is, by technical
reasons, limited to nuclei with $A<90\div100$.

For heavier nuclei, the challenge of experimentalists  was partially
responded within the self-consistent TFFS for magnetic moments
\cite{mu1,mu2} and quadrupole moments \cite{BE2,QEPJ,QEPJ-Web} of
odd spherical nuclei. Mention also the first self-consistent
calculation of the quadrupole moments of the $2^+_1$ states
\cite{Q2pl}. The calculations were mainly limited to semi-magic
nuclei considered in the ``single-quasiparticle approximation''
where one quasiparticle in the fixed state $\lambda=(n,l,j,m)$ with
the energy $\varepsilon_{\lambda}$ is added to the even-even core.
The QRPA-like TFFS equations for the nuclear response to the
external field, magnetic in \cite{mu1,mu2} or quadrupole in
\cite{BE2,QEPJ,QEPJ-Web}, were solved on the base of the Energy
Density Functional (EDF) by Fayans et al. \cite{Fay1,Fay5,Fay}. For
magnetic moments, the original Fayans functional DF3 \cite{Fay5,Fay}
was used whereas for quadrupole moments it was used together with
its modification, DF3-a, which was introduced in \cite{Tol-Sap} to
extend this approach to nuclei heavier than  lead. It differs from
the original one by the spin-orbit and effective tensor terms
which are important only for the fine structure of the
single-particle spectrum. For the quadrupole moments, the difference
between the predictions of the two functionals turned out noticeable
with a preference of DF3-a. We also mention that the same is true
for the energies and excitation probabilites of the $2^+_1$ states
in the lead and tin isotopes \cite{BE2,BE2-Web}.

 According to the TFFS \cite{AB1}, a quasiparticle differs from
a particle of the single-particle model in two respects. First, it
possesses a local charge $e_q$  and, second, the core is polarized
 due to the interaction between the particle and the core nucleons via the
 Landau--Migdal (LM) amplitude ${\cal F}$. In other words, the quasiparticle possesses the
 effective charge $e_{\rm eff}$ caused by the polarizability  of the core which is found
 by solving the TFFS equations.  In the MPSM,
 a similar quantity is  introduced  as a phenomenological parameter.
Thus, the consideration was made on the RPA (or QRPA) level which is
the ``zero approximation'' to the problem, corrections due to phonon
coupling (PC) effects were only estimated in \cite{mu1,mu2,QEPJ}.
This does not concern the investigation  of quadrupole moments of
the $2^+_1$ states in Ref. \cite{Q2pl} as the effect under
consideration there is beyond QRPA itself. On average, reasonable
description of the data was obtained in the Refs. cited above, with
the accuracy of $\delta \mu \simeq 0.1\div0.2\;$n.m. for magnetic
moments, and $\delta Q \simeq 0.1\div0.2\;$b for quadrupole moments.
This indicates that the single-quasiparticle approximation works
well for such nuclei on average, and the PC effects are usually
regular and  included in the TFFS parameters, i.e. the charges $e_q$
and the LM amplitudes ${\cal F}$. However, there are several ``bad''
cases, with $\delta \mu \simeq 0.5\;$n.m. for magnetic moments, and
$\delta Q \simeq 0.5\;$b for quadrupole moments which were
attributed to some non-regular PC effects. The estimations in
\cite{mu2} for magnetic moments and in \cite{QEPJ} for quadrupole
ones have shown that this interpretation looks reasonable and a more
detailed analysis of the PC effects is necessary.

In this context, it is worth citing older calculations of magnetic
moments within the conventional TFFS with the use of the Saxon-Woods
potential as the nuclear mean field \cite{Ba-Sp,TBK,BTF}.

Dealing with the PC problem within TFFS or other microscopic
approaches with fitted parameters, one faces  the problem of
refitting the parameters provided all the PC corrections are taken
into account. Exactly this way was chosen in \cite{PRL}, where a new
set parameters of the Skyrme functional was found which corrects for
 the PC effects. The same mode of action was
used in the first version of the self-consistent TFFS \cite{KhS}. In
this paper, dealing with magnetic moments of magic and semi-magic
nuclei, we chose a simpler way by separating only the kinds of the
PC diagrams which behave in non-regular way depending significantly
on the nucleus and the single-particle state under consideration.
The rest of the PC corrections, in agreement with the estimations,
is supposed to be regular and is included in the standard TFFS
parameters. Within this model, we analyzed first odd neighbors of
the  magic $^{208}$Pb nucleus. In this case the PC effects appear
mainly due to the $3^-_1$ state, although sometimes the sum of
contributions of all other phonons is also important. For these
nuclei, the problem was analyzed in the well-known paper by I.
Hamamoto \cite{Hamam} within the Nuclear Field Theory \cite{NFT},
the approach which operates with a set of phenomenological
parameters for each nucleus under consideration. On a microscopic
level, this problem was considered by Platonov \cite{Plat-mu} and
Tselyaev \cite{Tsel}. In the last article, some sets of higher order
terms in $g_L^2$ were summed up within QTBA.

In non-magic nuclei, the contribution of the $2^+_1$ state
dominates. Again we limit ourselves to the semi-magic nuclei where
this state is usually not too collective and the parameter $g_L^2$
is, as a rule, small. In addition, we will consider only such odd
semi-magic nuclei where the odd nucleon belongs to the normal
sub-system which simplifies the calculations because all the
equations for the PC corrections do not include pairing effects.
For example, in the lead region we will analyze the odd Tl and Bi
isotopes, but not the odd Pb nuclei, except $^{207,209}$Pb.

\section{Self-consistent TFFS}

For completeness, we outline first the TFFS formalism for the
effective magnetic field without phonon corrections. The magnetic
moment of an odd nucleus or the probability of the $M1$-transition
in such a nucleus are determined from the effective field $V$ which
is related to the external field $V_0=\hat {\bfg \mu}$, where $\hat
{\bfg \mu}$ is the operator of the magnetic moment: \beq \hat{\bfg
\mu} = g_l \hat{\bf l} + \frac{1}{2} g_s\hat{\bfg
\sigma},\label{mu0} \eeq with $g_l^p{=}1$, $g_l^n{=}0$,
$g_s^p{=}5.586$, and $g_s^n{=}-3.826$. The effective field $V$ obeys
an equation which is similar to the RPA equation with the LM
amplitude ${\cal F}$ playing the role of the effective
NN-interaction. In systems without pairing, e.g.  magic nuclei, this
equation reads \beq V(M1;\omega)= V_0(M1) + {\cal
F}A(\omega)V(M1;\omega), \label{Vef} \eeq where
$V_0(M1)=e_q\hat{\bfg{\mu}}$ and $A(\omega)=\int
G(\eps)G(\eps+\omega)d\eps/(2\pi i)$ stands for the particle-hole
propagator, $G(\eps)$ being the single-particle Green function. This
is illustrated diagrammatically on Fig. 1. The magnetic moments of
an odd nucleus with the odd nucleon in the one-quasiparticle state
$\lambda=(n_r,l,j,m)=(\nu,m)$ are the diagonal matrix elements of
the static effective field, $\mu_{\lambda}=\langle \lambda |
V(M1;\omega=0) |\lambda \rangle|_{m=j}$, whereas the matrix elements
of the $M1$-transition $\lambda_1 \to \lambda_2$  between two states
with the energies $\eps_1$ and $\eps_2$ respectively are given by
the non-diagonal matrix elements $M_{12}=\langle \lambda_1|
V(M1;\omega=\eps_2-\eps_1) |\lambda_2 \rangle$.

\begin{figure}[]
\vspace{10mm} \centerline {\includegraphics [width=60mm]{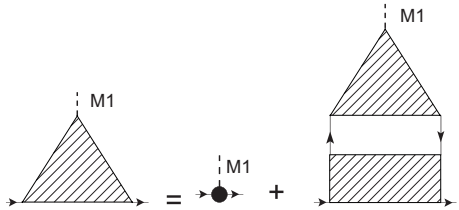}}
\vspace{2mm} \caption{ The effective magnetic field $V(M1)$. The
meaning of separate blocks becomes obvious after comparing with  Eq.
(\ref{Vef}). }
\end{figure}

In the explicit form,  the local quasiparticle charge $e_q$ in
Eq.~(\ref{Vef}) is defined as follows: \bea e_q\hat{\bfg\mu}&=&
\frac{1+(1-2\zeta_l)\hat{\tau_3}}2\,\hat{\bf l}\nonumber\\
&+&\frac{(g_s^p+g_s^n)+(g_s^p-g_s^n)
(1-2\zeta_s)\hat{\tau_3}}4\,\hat{\bfg\sigma}\nonumber\\
&+&\zeta_t[Y_2({\bf n})\circledast\hat{\bfg\sigma}]^1 .\label{eq}
\eea Here, following Refs.\cite{mu1,mu2}, in addition to the usual local
charge parameters $\zeta_s,\zeta_l$ of the FFS theory
\cite{AB1}, a new ``tensor'' (or ``$l$-forbidden'') charge
$\zeta_t$ is introduced. Terms of similar structure appear naturally
from MEC contributions to the magnetic moments \cite{Towner}.

In nuclei with pairing, Eq.~(\ref{Vef}) transforms to a set of three
equations \cite{AB1} \beq {\hat V}(\omega)={\hat V}_0(\omega)+{\hat
{\cal F}}  {\hat A}(\omega) {\hat V}(\omega), \label{Vef_s} \eeq
where all the terms are  matrices. In the standard TFFS notation
\cite{AB1}, we have: \beq {\hat V}=\left(\begin{array}{c}V
\\d_1\\d_2\end{array}\right)\,,\quad{\hat
V}_0=\left(\begin{array}{c}V_0
\\0\\0\end{array}\right)\,,
\label{Vs} \eeq where the fields $d_{1,2}$ denote variations of the
gap functions $\Delta^{(1),(2)}$ in the external field $V_0$. The
vector analogue of (\ref{Vef}) contains  $3\otimes3$ matrices: \beq
{\hat {\cal F}}=\left(\begin{array}{ccc}
{\cal F} &{\cal F}^{\omega \xi}&{\cal F}^{\omega \xi}\\
{\cal F}^{\xi \omega }&{\cal F}^\xi  &{\cal F}^{\xi \omega }\\
{\cal F}^{\xi \omega }&{\cal F}^{\xi \omega }& {\cal F}^\xi
\end{array}\right), \label{Fs} \eeq
and \beq {\hat A}(\omega)=\left(\begin{array}{ccc} {\cal L}(\omega)
&{\cal M}_1(\omega)
&{\cal M}_2(\omega)\\
 {\cal O}(\omega)&-{\cal N}_1(\omega) &{\cal N}_2(\omega)\\{\cal O}(-\omega)&-{\cal N}_1(-\omega) &
 {\cal N}_2(-\omega)
\end{array}\right)\,,
\label{As} \eeq where ${\cal L},\; {\cal M}_1$, and so on stand for
integrals over $\eps$ of the products of different combinations of
the Green function $G(\eps)$ and two Gor'kov functios
$F^{(1)}(\eps)$ and $F^{(2)}(\eps)$ \cite{AB1}.

In the matrix ${\hat {\cal F}}$, the diagonal terms  are the usual
LM amplitude ${\cal F}$ of the effective interaction in the
particle-hole (ph) and the interaction amplitude ${\cal F}^\xi$
irreducible in the pp (or hh) channel. The non-diagonal terms couple
the ph channel with the pp and hh ones.   As the analysis of
\cite{mu1} has shown, the interaction amplitude  ${\cal F}^\xi$ for
the $M1$ symmetry is, evidently, small. Therefore, we put ${\cal
F}^\xi{=}0$. Then, it is natural also to put equal to zero all the
non-diagonal terms of ${\cal F}^\xi$.  In this case, we have
$d_1=d_2=0$ and the only modification of Eq.~(\ref{Vef}) is the
change $A\to {\cal L}$,
 \beq {\cal
L(\omega)}=\int\frac{d\varepsilon}{2\pi
i}\left[G(\varepsilon)G(\varepsilon+\omega)-
F_1(\varepsilon)F_2(\varepsilon+\omega) \hat P \right]\,, \label{Ls}
\eeq where $\hat P=1$ for  T-even fields and $\hat P=-1$ for the
T-odd ones. To calculate the propagator ${\cal L}$ with consistent
account for  the continuum states, we use the generalization
\cite{PlSap} for superfluid
 systems of the Shlomo--Bertsch method \cite{ShB}.

All calculations below will be carried out with the use of the
self-consistent basis generated with the Generalized EDF by Fayans
et al.,  \beq E_0=\int {\cal E}[\rho_n({\bf r}),\rho_p({\bf
r}),\nu_n({\bf r}),\nu_p({\bf r})] d^3r,\label{E0} \eeq depending
simultaneously on the  normal $\rho_{\tau}$ and anomalous
$\nu_{\tau}$ densities. The DF3-a version of the normal EDF
\cite{Tol-Sap} will be used and the ``surface'' anomalous EDF
\cite{Fay}. More details can be found in \cite{BE2}.

For  magnetic moments we consider only the spin-dependent parameters
contribute.  The spin-dependent LM amplitude, just as in
\cite{mu1,mu2}, is chosen in the form of \beq {\cal F}^{\rm spin} =
{\cal F}_0^{\rm spin} + {\cal F}_{\pi} + {\cal F_\rho}, \label{Ftot}
\eeq where the pion and $\rho$-meson exchange terms are added to the
central force term ${\cal F}_0$. The central force was approximated
with the zero Landau harmonics. Following to \cite{scat_1,scat_2},
we take into account the momentum dependence of the amplitude $g'$
(``Migdal force''): \beq {\cal F}_0^{\rm spin}=C_0\left[
g\,({\bfg\sigma}_1{\bfg\sigma}_2)
+g'(q)({\bfg\sigma}_1{\bfg\sigma}_2) ({\bfg\tau}_1{\bfg\tau}_2)
\right] , \label{F0} \eeq
 \beq
g'(q)=\frac {g'}{1+r_0^2 q^2}, \label{g'k} \eeq where the
normalization factor $C_0=dn/d\eps_{\rm F})^{-1}=300\;$MeV$\cdot$
fm$^3$, and the value of $r_0=0.4\;$fm was found in
\cite{scat_1,scat_2} for the parameter which determines the
dependence of the amplitude $g'$ on the momentum transfer.

To evaluate the PC corrections we need in the vertex ${\hat
g_L}(r)$, the creation amplitude of the $L$-phonon. It  obeys the
homogeneous equation corresponding to Eq. (\ref{Vef_s}), \beq {\hat
g_L}(\omega)={\hat {\cal F}}  {\hat A}(\omega) {\hat g_L}(\omega),
\label{g_L} \eeq and is normalized as follows \cite{AB1},
\begin{equation}\label{norm}
\left({\hat g}_L^+ \frac {d {\hat A}}{d\omega}{\hat g}_L
\right)_{\omega=\omega_L}=-1,
\end{equation}
with the following notation:
\beq {\hat g_L}=\left(\begin{array}{c}g_L^{(0)}
\\g_L^{(1)}\\g_L^{(2)}\end{array}\right)\,,
\label{gL_s} \eeq

All the low-lying phonons we will consider have  natural parity. In
this case, the vertex $\hat g_L$ possesses  even $T$-parity. It is a
sum of two components with spins $S=0$ and $S=1$, respectively: \beq
\hat g_L= \hat g_{L0}(r) T_{LL0}({\bf n,\alpha}) + \hat g_{L1}(r)
T_{LL1}({\bf n,\alpha}), \label{gLS01} \eeq where $T_{JLS}$ stand
for the usual spin-angular tensor operators \cite{BM1}. The
operators $T_{LL0}$ and $T_{LL1}$ have  opposite $T$-parities, hence
the spin component should be the odd function of the excitation
energy, $g_L^{(1)}\propto \omega_L$.  In this case, the LM amplitude
in Eq. (\ref{g_L}) is also the sum, \beq {\cal F} = {\cal F}_0 +
{\cal F}^{\rm spin}, \label{F_tot} \eeq where the spin-independent
LM amplitude is generated by the generalized EDF in Eq.
(\ref{E0}),   \beq {\cal F}_0=\frac {\delta^2 {\cal E}}{\delta
\rho^2}. \label{LM0} \eeq The non-diagonal components of the matrix
(\ref{Fs}), the amplitudes ${\cal F}^{\omega \xi}={\cal F}^{\xi
\omega}$ stand for the mixed second derivatives, \beq {\cal
F}^{\omega \xi}=\frac {\delta^2 {\cal E}}{\delta \rho \delta \nu},
\label{LMxi} \eeq and, finally, the amplitude ${\cal F}^{\xi}$ is
the effective pairing interaction entering the gap equation, \beq
\Delta = {\cal F}^{\xi} \nu . \label{gap}\eeq The isotopic indices
in Eqs. (\ref{g_L} -- \ref{gap}) are for brevity omitted. In the
case of the surface pairing we deal, the ``mixed'' amplitude  ${\cal
F}^{\omega \xi}$ is rather important for the correct solution of the
self-consistent QRPA equation (\ref{g_L}).

\section{PC corrections in magic and semi-magic nuclei}

Let us now consider the PC corrections to Eq. (\ref{Vef}) induced by
a $L$-phonon. Keeping in mind the smallness of the $g_L^2$ parameter,
we limit ourselves to the so-called $g^2_L$-approximation. In
addition, as it was mentioned in the Introduction, we limit
ourselves with the case where the odd nucleon belongs to the
non-superfluid sub-system.  We will consider mainly the PC
corrections for this subsystem, therefore  the component $g_L^{(0)}$
of the vector (\ref{gL_s}) will participate only. To shorten
notation, we will omit for a time the upper index putting
$g_L^{(0)}\to g_L$. To be more exact, we deal with the PC
corrections to the matrix element $V_{\lambda_2
\lambda_1}=(\phi_{\lambda_2},V(M1) \phi_{\lambda_1})$ which is the
second order variation in the phonon field $g_L$ , see Fig. 2, \bea
\delta^{(2)} V_{\lambda_2 \lambda_1}&=&(\delta^{(2)}\phi_{\lambda_2},V
\phi_{\lambda_1})+ (\phi_{\lambda_2},V \;\delta^{(2)}\phi_{\lambda_1})\qquad
\qquad  \qquad  \qquad  \nonumber\\
&+& (\delta^{(1)}\phi_{\lambda_2},V \;\delta^{(1)}\phi_{\lambda_1}) {+}(\phi_{\lambda_2},
\delta^{(2)} V \phi_{\lambda_1}) \; \;\nonumber\\
&+& (\delta^{(1)}\phi_{\lambda_2},\delta^{(1)} V \phi_{\lambda_1}) +
(\phi_{\lambda_2},\delta^{(1)}V \;\delta^{(1)}
\phi_{\lambda_1}),  \label{cor-gen} \eea with the obvious
notation. In the cases where it can not lead to misleading, we shorten the notation below as
follows $\delta^{(2)} \to \delta$.

\begin{figure}[]
\vspace{2mm} \centerline {\includegraphics [width=30mm]{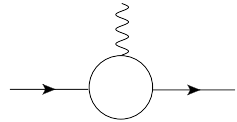}}
\vspace{2mm} \caption{  The  $L$-phonon creation amplitude $g_L({\bf
r})$. }
\end{figure}

\subsection{PC corrections included into the model}
We  separate those terms of Eq. (\ref{cor-gen}) which may behave in
non-regular way changing significantly depending on the nucleus and
the state under considerations. As it will be argued in the next
subsection, this does concern the three first  terms of Eq.
(\ref{cor-gen}).  The last two terms disposed in the third line are regular and can
be included into the TFFS parameters. As to the forth term, $\delta^{(2)} V$, it contains both, the
regular and non-regular contributions.

The terms in the first line of (\ref{cor-gen}), the second order
variations of the single-particle wave functions may be named as the
``end correction'', keeping in mind the ends of the diagram of Fig.
1 for the effective field.  The  main, pole part of the end
correction corresponding to the second of these two terms is
illustrated  in Fig. 3. The corresponding tadpole term is shown in
Fig. 4. The phonon $D$-function appears here and below after connecting  two wavy ends
of Fig. 2 which corresponds to averaging of the product of two boson
(phonon) operators $B^+B$ over the ground state of the nucleus without
phonons.

\begin{figure}[]
\vspace{10mm} \centerline {\includegraphics [width=80mm]{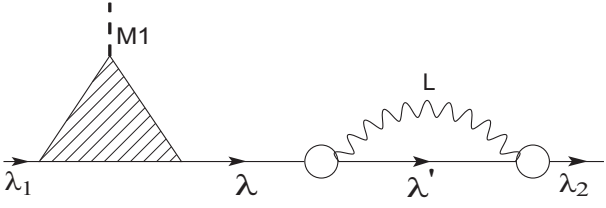}}
\vspace{2mm} \caption{ A correction to the ``end''. The open blob is
the $L$-phonon creation amplitude  $g_L({\bf r})$. The wavy line
denotes the phonon $D$-function.}
\end{figure}

\begin{figure}[]
\vspace{10mm} \centerline {\includegraphics [width=80mm]{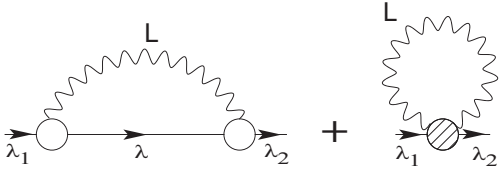}}
\vspace{2mm} \caption{PC corrections to the mass operator. The
dashed blob denotes the ``tadpole'' term.}
\end{figure}

For the matrix element $V_{\lambda_2\lambda_1}$   of the effective
field, the sum of these corrections  corresponds to the following
formula:
\bea \delta V_{\lambda_2\lambda_1}^{\rm end} & = & -
\sum_{\lambda}V_{\lambda_2\lambda}G_{\lambda}(\eps_{\lambda_1}) \delta
\Sigma_{\lambda\lambda_1}(\eps_{\lambda_1})
   \nonumber \\
& - & \sum_{\lambda}\delta
\Sigma_{\lambda_2\lambda}(\eps_{\lambda_2})
G_{\lambda}(\eps_{\lambda_2}) V_{\lambda\lambda_1}\label{end1}.\eea
Here $\delta \Sigma$ means the PC correction to the mass operator
$\Sigma$ displayed in Fig. 4. In addition to the pole diagram of
Fig. 4, we showed also for completeness the non-pole ``tadpole'' one
\cite{KhS}. As we will see, this term $\delta\Sigma^{\rm tad}$ does
practically not contribute to the problem under consideration.

The pole diagram in Fig. 4 corresponds to the following expression,
\bea \delta\Sigma_{\lambda_2\lambda_1}(\eps) &=& \int
\frac{d\omega}{2\pi i} \sum_{\lambda\,M}<\lambda_2|g^+_{LM}|\lambda>
<\lambda|g_{LM}|\lambda_1> \nonumber\\
&\times& D_L(\omega)G_{\lambda}(\eps-\omega), \label{dSig1} \eea
where $D_L(\omega)$ is the phonon $D$-function.  After an
integration, one obtains \bea
\delta\Sigma_{\lambda_2\lambda_1}(\epsilon)&=&\sum_{\lambda\,M}<\lambda_2|g^+_{LM}|\lambda>
<\lambda|g_{LM}|\lambda_1> \nonumber\\
&\times&\left(\frac{n_{\lambda}}{\eps+\omega_L-
\eps_{\lambda}}+\frac{1-n_{\lambda}}{\eps-\omega_L
-\eps_{\lambda}}\right), \label{dSig2} \eea
where $\omega_L$ is the
excitation energy of the $L$-phonon and $n_{\lambda}=(0,1)$ stands
for the occupation numbers.

After substitution of (\ref{dSig2}) into (\ref{end1})  one obtains:
\bea  \delta V_{\lambda_2\lambda_1}^{\rm end}& =& - \sum_{\lambda'
\lambda M} \frac { V_{\lambda_2\lambda}<\lambda|g^+_{LM}|\lambda'>
<\lambda'|g_{LM}|\lambda_1>  } {\eps_{\lambda_1} - \eps_{\lambda}}
 \nonumber\\ &\times&
 \left(\frac{n_{\lambda'}}{\eps_{\lambda_1}+\omega_L-
\eps_{\lambda'}}+\frac{1-n_{\lambda'}}{\eps_{\lambda_1}-\omega_L
-\eps_{\lambda'}}\right) \nonumber\\ & - &
\sum_{\lambda \lambda' M}\left(\frac{n_{\lambda'}}{\eps_{\lambda_2}+\omega_L-
\eps_{\lambda'}}+\frac{1-n_{\lambda'}}{\eps_{\lambda_2}-\omega_L
-\eps_{\lambda'}}\right)
 \nonumber\\ &\times&\frac { <\lambda_2|g^+_{LM}|\lambda'>
<\lambda'|g_{LM}|\lambda>V_{\lambda\lambda_1} } {\eps_{\lambda_2} - \eps_{\lambda}}.\label{end2}  \eea

In this equation, the term with $\lambda=\lambda_1$  in the first sum and the one with $\lambda=\lambda_2$
in the second sum  are singular.
This singularity is removed with the standard renormalization \cite{KhS} of the single particle
wave functions $\varphi_{\lambda_1}\to \sqrt{Z_{\lambda_1}} \varphi_{\lambda_1}$,
$\varphi_{\lambda_2}\to \sqrt{Z_{\lambda_2}} \varphi_{\lambda_2}$, where \beq Z_{\lambda} =
\left( 1- \left. \frac {\partial
\Sigma_{\lambda\lambda}(\eps)}{\partial
\eps}\right|_{\eps=\eps_{\lambda}}\right)^{-1} \label{Zlam} \eeq is
the residue of the Green function at the pole $\eps=\eps_{\lambda}$.

To avoid misleading, we note that below we will for brevity name
``$Z$-factor'' not the total quantity but the PC contribution only.
Correspondingly, we insert into Eq. (\ref{Zlam}) the PC term $\delta
\Sigma$ only. In the TFFS the total $Z$-factor enters into the LM
amplitude ${\cal F}=Z^2 \Gamma^{\omega}$, the local charges
$e_q=Z{\cal T}^{\omega}$ and the mean field $U(r)=Z\Sigma_0(r)$,
where the subscript ``0'' means that the energy and momentum  are
taken at the Fermi surface.  Here the notation  \cite{AB1} is used
with the only change $a\to Z$. The average value of the $Z$-factor
corresponding to the total mass operator $\Sigma_0$ was estimated in
\cite{KhS} and \cite{Z-fac1,Z-fac2} as $Z_0=0.8$.

The rest of these sums with non-diagonal terms $\lambda \neq
\lambda_1$ in the first case and $\lambda \neq \lambda_2$ in the
second can be calculated directly and, as we shall see, are rather
small. However, we retain them for completeness, and represent the
``end correction'' as the sum: \beq  \delta
V_{\lambda_2\lambda_1}^{\rm end} = \delta V_{\lambda_2\lambda_1}^Z +
(\delta V_{\lambda_2\lambda_1}^{\rm end})',   \label{end3}  \eeq
where \beq \delta{V}^Z_{\lambda_2\lambda_1} =
\left(\sqrt{Z_{\lambda_2}Z_{\lambda_1}}-1\right)
V_{\lambda_2\lambda_1}. \label{endZ} \eeq Eqs. (\ref{end3}),
(\ref{endZ}) correspond
 to partial summation of the diagrams of Fig. 3, and, hence, contains  higher order  terms in $g_L^2$.
 To be consistent up to the order $g_L^2$, the $Z$-factors
Eq. (\ref{endZ}) should be expanded in terms of $\partial
\Sigma_{\lambda\lambda}(\eps) /
\partial \eps$, $Z=1+ \partial \Sigma_{\lambda\lambda}(\eps) /
\partial \eps$, with the result
\beq \delta{V}^Z_{\lambda_2\lambda_1} {=} \frac 1 2 \left( \left.
\frac  {\partial \Sigma_{\lambda_2\lambda_2}(\eps)}{\partial
\eps}\right|_{\eps=\eps_{\lambda_2}} {+} \left. \frac  {\partial
\Sigma_{\lambda_1\lambda_1}(\eps)}{\partial
\eps}\right|_{\eps=\eps_{\lambda_1}} \right) V_{\lambda_2\lambda_1}.
\label{endZ1} \eeq This contribution is important for the
conservation of the total angular momentum. This will be shown
below.

As the term $\delta\Sigma^{\rm tad}$ does not depend on the energy
$\eps$ it does not contribute to $Z_{\lambda}$  and, hence, to the
main term representing the end effect which is given by Eq.
(\ref{endZ}). It  may contribute to the second term of the sum of
(\ref{end3}) which itself is negligibly small.

The energy derivative of the mass operator (\ref{dSig2}) is as
follows: \bea \left.
\frac{\partial\delta\Sigma_{\lambda\lambda}(\eps)}{\partial\epsilon}\right|_{\eps=\eps_{\lambda}}
&=&-\sum_{\lambda'\,M}|<\lambda'|g_{LM}|\lambda>|^2 \nonumber\\
\times \left[\frac{n_{\lambda'}}{(\eps_{\lambda}+\omega_L-
\eps_{\lambda'})^2}\right. &+&\left. \frac{1-n_{\lambda'}}
{(\eps_{\lambda} -\omega_L -\eps_{\lambda'})^2} \right].
\label{ddSig} \eea

\begin{figure}[]
\vspace{10mm} \centerline {\includegraphics [width=40mm]{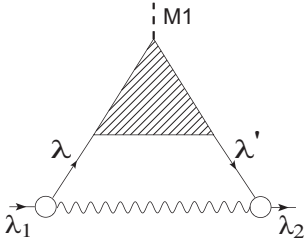}}
\vspace{2mm} \caption{ The correction due to the induced
interaction.}
\end{figure}

Let us now go to the second line of Eq. (\ref{cor-gen}) and begin with the
first term containing the first variations of the wave functions. After connecting two wavy ends
we obtain the ``triangle'' diagram
($GGD$) displayed in Fig. 5 which describes the effect of the induced
interaction ${\cal F}_{\rm ind}$ due to exchange of the $L$-phonon.
The explicit form of the corresponding correction to the matrix
element $V_{\lambda_2\lambda_1}$ is \bea \delta
V_{\lambda_2\lambda_1}^{GGD}  &= & \int \frac {d\omega}{2\pi i}
\sum_{\lambda\,\lambda'\,M}<\lambda_2|g^+_{LM}|\lambda'>V_{\lambda'\lambda}
<\lambda|g_{LM}|\lambda_1> \nonumber\\
&\times&
G_{\lambda}(\eps_{\lambda_1}-\omega)G_{\lambda'}(\eps_{\lambda_2}-\omega)
D_L(\omega). \label{GGD} \eea After separating the angular variables
and integrating over the energy, we obtain:

\beq \delta V_{\lambda_2\lambda_1}^{GGD} =(-1)^{j_2-m_2}
\!\left(\!\begin{array}{ccc} j_2&1& j_1\\-m_2& M&
m_1\end{array}\right)\!<\!2 \!\parallel \delta V^{GGD} \parallel\!
1\!>,\label{GGD1}\eeq where the short notation $1=\nu_1$ is used and
the reduced matrix element is equal to \bea <2 \parallel \delta
V^{GGD} \parallel 1>  = \sum_{34} (-1)^{L+j_4-j_3}
\left\{\!\begin{array}{ccc} 1&j_1& j_2\\L&
j_4&j_3\end{array}\right\}
  \nonumber\\
 \times <4\!\parallel V \parallel\! 3 >
 <2\!\parallel \tilde g_L \parallel\! 4 > <3\!\parallel g_L \parallel\! 1> I_{34}(\omega_L),
 \quad \label{GGD2}\eea
 \bea I_{34}(\omega_L)= \frac 1 {(\eps_3-\eps_4)-(\eps_1-\eps_2)} \left[\frac {n_3}
{\eps_1-\eps_3+\omega_L}\right.\nonumber\\
\left.  - \frac {n_4} {\eps_2-\eps_4+\omega_L}
 {+}\frac {1-n_3}
{\eps_1-\eps_3-\omega_L} {-} \frac {1-n_4} {\eps_2-\eps_4-\omega_L}
\right], \label{GGD3}\eea where \bea  \tilde g_L=g_L(-\omega)=
g_{L0}(r;\omega) T_{LL0}({\bf n,\alpha}) \nonumber\\ -
g_{L1}(r;\omega) T_{LL1}({\bf n,\alpha}), \label{gLtild} \eea

\begin{figure}[]
\vspace{10mm} \centerline {\includegraphics [width=40mm]{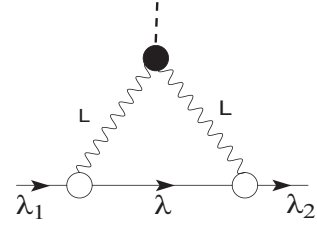}}
\vspace{2mm} \caption{ Correction due to the direct action of the
$M1$-field to the phonon. The black blob is the magnetic moment of
the phonon.}
\end{figure}

Let us go to the second term in the second line of Eq. (\ref{cor-gen})
with the second order variation of the effective field $V$. This term was
considered in detail in \cite{Kaev2011}.
We separate the PC correction  due to the
phonon contribution to the local charge $e_q$ in the first term of
Eq. (\ref{Vef}) for the effective field.  It is represented by the
diagram of Fig. 6 describing the contribution of the phonon magnetic
moment. It can be obtained from the first diagram of Fig. 4 by
inserting the external magnetic field in the phonon line.
 In Fig. 6, the black blob means the phonon magnetic
moment, \beq \mu_L^{\rm ph}=g_L^{\rm ph} L, \label{mu-ph} \eeq where
$g_L^{\rm ph} $ is the phonon gyromagnetic ratio. In the
Bohr--Mottelson model \cite{BM2,Hamam}, one has $g^{\rm ph}_{L,\rm
BM}=Z/A$. Similar to to Eq. (\ref{GGD}), we get \bea \delta
V_{\lambda_2\lambda_1}^{GDD} &= & \int \frac {d\omega}{2\pi i}
\sum_{\lambda\,\,M}<\lambda_2|g^+_{LM}|\lambda>
<\lambda|g_{LM}|\lambda_1> \nonumber\\
&\times& G_{\lambda}(\eps_{\lambda_1}-\omega) D_L(\omega)
D_L(\omega-\omega_0). \label{GDD} \eea

Again, after separating the angular variables, we obtain for the
reduced matrix element, \bea <2 \parallel \delta V^{GDD}
\parallel 1>  = \sum_3 (-1)^{L+j_1+j_3} \gamma_L \qquad\qquad \nonumber\\
 \times  \sqrt{ \frac {L(L+1)(2L+1)}{4\pi}}
\left\{\!\begin{array}{ccc} j_1&1& j_2\\L& j_3& L\end{array}\right\}
<\!3\!\parallel g_L \parallel \!1\!>
\nonumber\\
 \times    <\!2\! \parallel \tilde g_L \parallel\! 3\!>  \left(I_3^{(1)}
 (\omega_L)+I_3^{(2)}(\omega_L) \right),
  \qquad\qquad \label{GDD2}\eea
\bea I_3^{(1)}(\omega_L) {=} \frac
{1-n_3}{(\eps_1-\eps_3-\omega_L)(\eps_2-\eps_3
-\omega_L)} \nonumber\\
 \frac {n_3}{(\eps_1-\eps_3+\omega_L)(\eps_2-
\eps_3+\omega_L)}, \label{GDD3}\eea \bea I_3^{(2)}(\omega_L) {=}
\frac 1 {\eps_1{-}\eps_2{-}2\omega_L} \left( \frac
{n_3}{\eps_2{-}\eps_3{+}\omega_L} + \frac {1-n_3}
{\eps_1{-}\eps_3{-}\omega_L} \right)\qquad \nonumber\\
-\frac 1 {\eps_1{-}\eps_2{+}2\omega_L} \left( \frac
{1-n_3}{\eps_2-\eps_3-\omega_L} + \frac
{n_3}{\eps_1-\eps_3+\omega_L} \right). \quad \label{GDD4}\eea The
second integral (\ref{GDD4}) reveals a dangerous behavior at
$(\omega_L-\omega_0) \to 0$. For the case of the static external
field we deal, it occurs at  $\omega_L \to 0$.

\begin{figure}[]
\vspace{10mm} \centerline {\includegraphics [width=20mm]{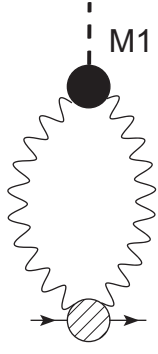}}
\vspace{2mm} \caption{ The tadpole-like diagram for the contribution
of the phonon magnetic moment.}
\end{figure}

Analogically to producing Fig. 6 by inserting the external field to
the first, pole, diagram of Fig. 4, there is the tadpole-like PC
correction to the effective magnetic field which corresponds to
inserting the external field to the phonon $D$-function in the
second diagram on Fig. 4. It is displayed in Fig. 7. The dashed blob
denotes the sum of all non-pole diagrams of phonon-particle
scattering  (the phonon-particle Compton diagrams). It contains the
integral \beq I^{\rm tad}= \int \frac {d\omega}{2\pi i} D_L(\omega)
D_L(\omega-\omega_0) = \frac {4\omega_L}{\omega_L^2-\omega_0^2}.
\label{tad6}\eeq Its behavior at $\omega_L \to 0$ is just the same
as of the integral $I_3^{(2)}$, Eq. (\ref{GDD4}). This makes it
reasonable to suppose that their sum is regular at
$(\omega_L-\omega_0) \to 0$.

The phonon gyromagnetic ratio $g_L^{\rm ph}$ for magic nuclei  is
obtained from the diagrams shown in Fig. 8. For non-magic nuclei,
the corresponding set of diagrams is much more complicated, see Ref.
\cite{Q2pl,Kaev2011} for the case of the quadrupole moment, and in
this case we limit ourselves to the BM model.

\begin{figure}[]
\vspace{10mm} \centerline {\includegraphics [width=80mm]{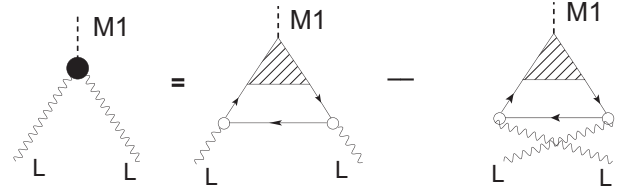}}
\vspace{2mm} \caption{ Diagrams for the phonon magnetic moment in
magic nuclei.}
\end{figure}

Thus, the sum of the PC corrections to the effective $M1$-field is
as follows: \bea \delta V &=& \delta V^Z + \delta V_{GGD} + \delta
V^{(1)}_{GDD}+ \delta V'_{\rm end}\nonumber\\ &+& \left[ \delta
V^{(2)}_{GDD} + \delta V_{\rm non-pole} \right], \label{PC-sum} \eea
where $\delta V^{(1),(2)}_{GDD}$ are the terms of Eq. (\ref{GDD})
with the integrals $I^{(1),(2)}$ and $\delta V_{\rm non-pole}$
denotes the sum of all non-pole PC corrections to the magnetic
effective field. Up to now, the method of a consistent calculation
of this quantity is implemented for magic nuclei mainly
\cite{KhS,Plat-mu}. For non-magic nuclei, the formalism was
developed in \cite{Kam-Sap} but the only one realization was carried
out \cite{Q2pl}. As we shall see below, the magnitude of the first
two terms of the sum (\ref{PC-sum}) is, as a rule, significantly
larger than that of $\delta V^{(1),(2)}_{GDD}$ terms. However,
$\delta V^Z$ and $\delta V_{GGD}$ terms are always of the opposite
sign and strongly compensate each other. Therefore, the term $\delta
V^{(1)}_{GDD}$ is sometimes not negligible and should be taken into
account. In the case when the external field is $V_0={\bf j}$, where
${\bf j} = {\bf l}+ 1/2 {\bfg \sigma}$ is the total angular
momentum, due to the conservation of ${\bf j}$ condition, the total
PC correction (\ref{PC-sum}) should be zero. As we will discuss in
the next section, the sum of the first four terms of (\ref{PC-sum})
is equal to zero. This means that the terms $\delta V^{(2)}_{GDD}$
and $\delta V_{\rm non-pole}$ should compensate each other.
 We will suppose that a similar relation is
approximately valid for any $M1$-field and will use the model where
the terms in the first line of Eq. (\ref{PC-sum})  are taken into
account only   and those on the second line are neglected. In fact,
we carry out the summation of the end corrections, similar to that
in Eq. (\ref{endZ}), and the final ansatz we use is as follows: \bea
\widetilde{V}_{\lambda_2\lambda_1} &=&
\sqrt{Z_{\lambda_2}Z_{\lambda_1}} \nonumber \\
&\times&\left( V + \delta V_{GGD} + \delta V^{(1)}_{GDD} + \delta
V'_{\rm end}\right)_{\lambda_2\lambda_1}. \label{final} \eea Just as
Eq. (\ref{endZ}) {\it vs} Eqs. (\ref{end1}) and (\ref{end2}), the
ansatz (\ref{final}) differs from the prescription of Eq.
(\ref{PC-sum}) in terms higher in $g_L^2$.

In conclusion of this section, we write down the expression for the
$L$-phonon magnetic moment $\mu_L$ in magic nuclei corresponding to
Fig. 8:
\bea \mu_L {=} \sum_{123} ({-}1)^{L{+}1}
\begin{pmatrix}
1 &L &L\\
0 &L &-L
\end{pmatrix}
\left\lbrace\begin{matrix}
1 &L &L\\
j_3 &j_2 &j_1
\end{matrix}\right\rbrace  \qquad  \qquad \nonumber\\
 \times <1 \parallel V(M1) \parallel 2>   \left[<1\parallel g_L \parallel 3>
  <3\parallel \tilde g_L \parallel 2> \right.  \nonumber\\
{\times} I^{GGG}_{123}(\omega_L) {-} \left. <\!1\!\parallel \tilde g_L \parallel\!3\!>
  <\!3\!\parallel g_L \parallel\!2\!>I^{GGG}_{123}(-\omega_L) \right],\;\;
\label{GGG}\eea
where \bea I^{GGG}_{123}(\omega_L)&=&\frac 1 {\eps_2
-\eps_1} \left[ \frac
{n_1 (1-n_2) (1-n_3) {-}(1{-}n_1) n_2 n_3}
{\eps_1-\eps_3-\omega_L} \right.
\nonumber\\
& + &\left. \frac
{n_1 (1-n_2) n_3-(1-n_1) n_2
(1-n_3)}{\eps_2- \eps_3-\omega_L}
\right] \nonumber\\
&+&\frac {n_1 n_2(1-n_3) - (1-n_1)(1-n_2) n_3}
{(\eps_1-\eps_3-\omega_L)(\eps_2-\eps_3-\omega_L)}.
\label{GGG1} \eea

For the general case of $\omega _0\neq 0$,   $\omega_ L' = \omega_ L + \omega_0 \neq \omega_ L$
formulae analogous to Eqs. (\ref{GGG}) and (\ref{GGG1}) for the E2 symmetry have been obtained
 in Ref. \cite{KaV2011}.

\subsection{PC corrections we omit}

The diagram in  Fig. 6 corresponding the variation of the local
charge $e_q$ in the effective field $V$ is the only one example of
$g_L^2$ PC corrections to the effective field $\delta^{(2)}V$. The
main part of such corrections appears from variations of the
integral term of Eq. (\ref{Vef}).  These terms were examined in
detail in Refs. \cite{Kaev2011,Kaev83,Tsel}. The term corresponding
to the variation of the particle-hole propagator,
$\delta^{(2)}A{(1)}=\delta^{(1)}G  \delta^{(1)} G$ is displayed in
Fig. 9. The wavy line for the phonon $D$-function, just as above,
appears after connecting the phonon ends of Fig. 2.  This diagram,
together with that in Fig. 5, can be interpreted as the substitution
of the sum ${\cal F}^{\rm spin}\to {\cal F}^{\rm spin}+{\cal F}_{\rm
ind}$ as the effective interaction in Eq. (\ref{Vef}) with different
order of these operators. Another variation of the particle-hole
propagator, $\delta^{(2)}A{(2)}=2\delta^{(2)}G G$,  is displayed in
Fig. 10. There are also more complicated diagrams for
$\delta^{(2)}V$ containing mixed variations $\delta^{(1)} {\cal
F}^{\rm spin} \delta^{(1)}A$, see \cite{Kaev2011}.

\begin{figure}[]
\vspace{10mm} \centerline {\includegraphics [width=50mm]{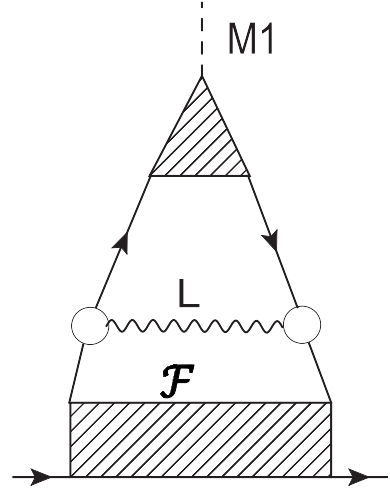}}
\vspace{2mm} \caption{The diagram alternative to the one in Fig. 5.}
\end{figure}

\begin{figure}[]
\vspace{10mm} \centerline {\includegraphics [width=50mm]{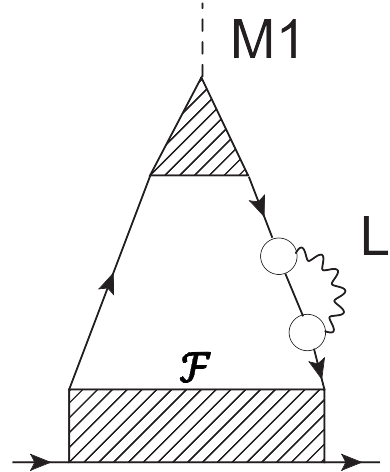}}
\vspace{2mm} \caption{A PC correction to the particle-hole
propagator in Eq. (\ref{Vef}).}
\end{figure}

Let us discuss the reasons why we exclude these two diagrams from
our model although they are of the same $g_L^2$ order. In fact, as
it is seen from all the equations of the above subsection, $g_L^2$
appears always with some energy dependent denominators, and the
characteristic dimensionless combination is the ratio of \beq
\beta_L = \frac {(g_L)_{\lambda \lambda'}^2}
{(\eps_{\lambda}-\eps_{\lambda'}\pm \omega_L)^2}. \label{betL}\eeq

Typical value of the matrix element $(g_L)_{\lambda \lambda'}\simeq
1\;$MeV whereas the energy differences in the denominator of
(\ref{betL}) are of the order of $\eps_{\rm F}/A^{1/3}\simeq 7\div
10\;$MeV. It should be reminded  that we deal with ``magic''
sub-system of a semi-magic nucleus where pairing is absent.
Therefore, as a rule, $\beta_L$ is really a small parameter and the
``$g^2_L$ expansion''  is justified, each separate term is small and
only a large sum of such contributions can be important. Such sums
do practically not depend on the nucleus under consideration and on
the single-particle state of the odd nucleon. As simple estimations
show,  this is true for diagrams in Fig. 9 and Fig. 10. According to
the strategy of our model, we must consider them as contributions to
the LM amplitude ${\cal F}^{\rm spin}$ and the local charge $e_q$
and not calculate explicitly. On the other hand, as we shall see
below, the PC corrections considered in the previous Subsection may
depend, sometimes strongly, on the state $\lambda_0$ of the odd
nucleon with rather big contribution of a separate term with small
energy denominator in $\beta_L$. Such PC corrections depend
significantly on the nucleus under consideration and on the state
$\lambda_0$ of the odd nucleon. They should be calculated
explicitly.

Fig. 11 represents an example of the ``mixed'' PC corrections from
the third line of Eq. (\ref{cor-gen}). All arguments to exclude the
 two diagrams from our model remain valid for them too.

\begin{figure}[]
\vspace{10mm} \centerline {\includegraphics [width=50mm]{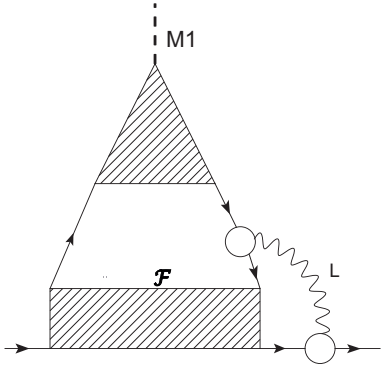}}
\vspace{2mm} \caption{An example of more complicated ``mixed''
diagrams corresponding to the third line of Eq. (\ref{cor-gen}).}
\end{figure}

\section{Phonon magnetic moments in $^{208}$Pb}
In this double-magic nucleus  several low-lying phonons are known
with varying degree of collectivity. The lowest one $3^-$ possesses
 the highest collectivity whereas the strength of the $5^-$ mode is
shared between two states. Before starting the study of PC effects
in $^{208}$Pb we analyze  the accuracy of describing the phonons
themselves. Their characteristics are presented in Table I. We see
that the most collective $3^-$ level is described sufficiently well,
$5^-_1$ and $5^-_2$ ones, a little worse. As to the $2^+_1$ state
and other states of positive parity, they are not strongly
collective as there is no low-energy particle-hole configurations of
positive parity except spin-orbit doublets $(h_{11/2}^{-1} h_{9/2})$
for protons and $(i_{13/2}^{-1} i_{11/2})$ for neutrons. In such a
situation, $\omega_L$ values in the QRPA solution are shifted only
little from the nearest particle-hole excitation energy,
$\eps_p(h_{9/2})-\eps_p(h_{11/2})$ or
$\eps_n(i_{11/2})-\eps_n(i_{13/2})$ in our case. Single-particle
levels in $^{208}$Pb generated with the DF3-a functional we use are
displayed in Fig. 12 for neutrons and Fig. 13 for protons.
Comparison is made with the experimental spectra and those obtained
by us with the one of the best Skyrme functionals HFB-17
\cite{HFB-17}. We see that our spectra reasonably  agree with the
data, on average, better than the HFB-17 one. However, some
inaccuracy takes place for levels under consideration which is
partially responsible for the shift of the $2^+_1$ level up to 660
keV. Another reason for this discrepancy is evidently not taking
into account the spin-orbit LM amplitude in Eq. (\ref{g_L}) for the
$g_L$ vertex. We shall see below that the PC corrections in magic
nuclei originate mainly from the $3^-$ phonon, therefore some
inaccuracy in describing higher collective states does not lead to
serious errors.

\begin{figure}[]
\vspace{10mm} \centerline {\includegraphics [width=70mm]{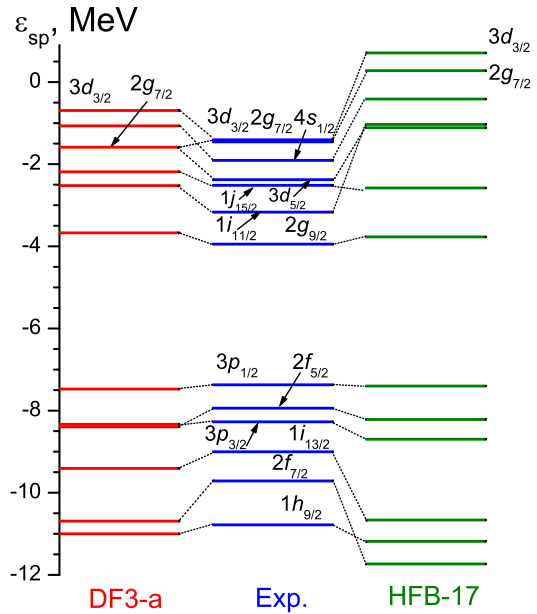}}
\vspace{2mm} \caption{ Neutron single-particle levels in
$^{208}$Pb.}
\end{figure}

\begin{figure}[]
\vspace{10mm} \centerline {\includegraphics [width=70mm]{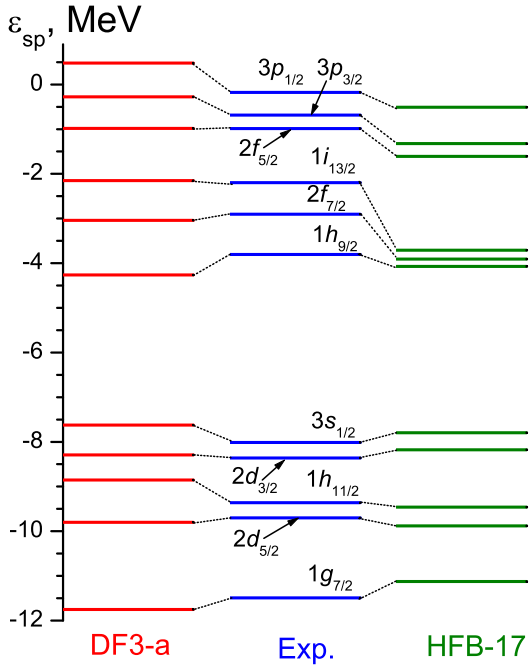}}
\vspace{2mm} \caption{ Proton single-particle levels in $^{208}$Pb.}
\end{figure}

\begin{table}[]
\caption{Characteristics of the low-lying phonons in $^{208}$Pb,
$\omega_L$ (MeV) and $B(EL,{\rm up)}$(${\rm e^2 fm}^{2L}$).}
\begin{tabular}{c c c c c  }

\hline \hline\noalign{\smallskip} $L^\pi$  & $\omega_L^{\rm th}$   &
$\omega_L^{\rm exp}$
&  $B(EL)^{\rm th}$ &  $B(EL)^{\rm exp}$  \\
\noalign{\smallskip}\hline\noalign{\smallskip}
$3^-_1$    & 2.684    &  2.615     &$7.093\times 10^5 $    & $6.12 \times 10^5 $\\
$5^-_1$    & 3.353    &  3.198     &$3.003\times 10^8 $    & $4.47 \times 10^8 $\\
$5^-_2$    & 3.787    &  3.708     &$1.785 \times 10^8 $   & $2.41 \times 10^8 $\\
$2^+_1$    & 4.747    &  4.086     &$1.886 \times 10^3 $   & $3.18 \times 10^3 $\\
$2^+_2$    & 5.004    &  4.928     &$1.148 \times 10^3 $   & - \\
$4^+_1$    & 4.716    &  4.324     &$3.007 \times 10^6 $   & - \\
$4^+_2$    & 5.367    &  4.911(?)  &$8.462 \times 10^6 $   & - \\
$6^+_1$    & 4.735    &   -        &$6.082 \times 10^9 $   & - \\
$6^+_2$    & 5.429    &   -        &$1.744 \times 10^{10}$ & - \\
\noalign{\smallskip}\hline \hline

\end{tabular}\label{tab1}
\end{table}

\begin{table*}[]
\caption{Magnetic moments (in $\mu_N$ units) of phonons in
 $^{208}$Pb.}
\begin{tabular}{c c c c c c c c c c c}

\hline \hline\noalign{\smallskip}

$L^{\pi}$ &$\mu_n^{(j)}$ &$\mu_n^{(s)}$  & $\mu_n$  & $\mu_p^{(j)}$&
$\mu_p^{(s)}$ & $\mu_p$ &
$\mu_L^{(j)}$ &$\mu_L^{(s)}$&$\mu_L$&$g_L^{\rm ph}$\\
\noalign{\smallskip}\hline\noalign{\smallskip} $3^-$\,\,&\,\,-0.074
\,\,&\,\,-0.039\,\,&\,\,-0.113\,\,&\,\,1.566\,\,&\,\,0.058\,\,&\,\,1.492\,\,
&\,\,1.492\,\,&\,\,0.019\,\,&\,\,1.511\,\,&\,\,0.463\\

$5^-_1$ &-0.027 &-0.018 &-0.046   & 4.733 & 0.278 & 5.011 & 4.705 & 0.260 & 4.965  & 0.993  \\

$5^-_2$ &-0.215 &-0.478 &-0.693   & 0.853 &-0.123 & 0.730 & 0.638 &-0.600  & 0.037  & 0.008  \\

$2^+_1$ &-0.027 & 0.000 &-0.027   & 1.536 & 0.493 & 2.029 & 1.509 & 0.492 & 2.002  & 1.001   \\

$2^+_2$ &-0.027 & 0.004 &-0.022   & 1.541 & 0.406 & 1.947 & 1.514 & 0.411 & 1.925  & 0.962   \\

$4^+_1$ &-0.009 &-0.010 &-0.018   & 4.017 & 0.449 & 4.466 & 4.008 & 0.440 & 4.448  & 1.112   \\

$4^+_2$ &-0.112 &-0.232 &-0.343   & 1.822 & 0.276 & 2.098 & 1.711 & 0.044 & 1.755  & 0.439   \\

$6^+_1$ &-0.005 &-0.004 &-0.009   & 6.172 & 0.294 & 6.466 & 6.167 & 0.290 & 6.457  & 1.076   \\

$6^+_2$ &-0.075 &-0.147 &-0.222   & 4.765 & 0.092 & 4.857 & 4.690 &-0.054 & 4.636  & 0.773   \\

\noalign{\smallskip}\hline \hline
\end{tabular}\label{tab2}
\end{table*}

\begin{figure}[]
\vspace{10mm} \centerline {\includegraphics [width=86mm]{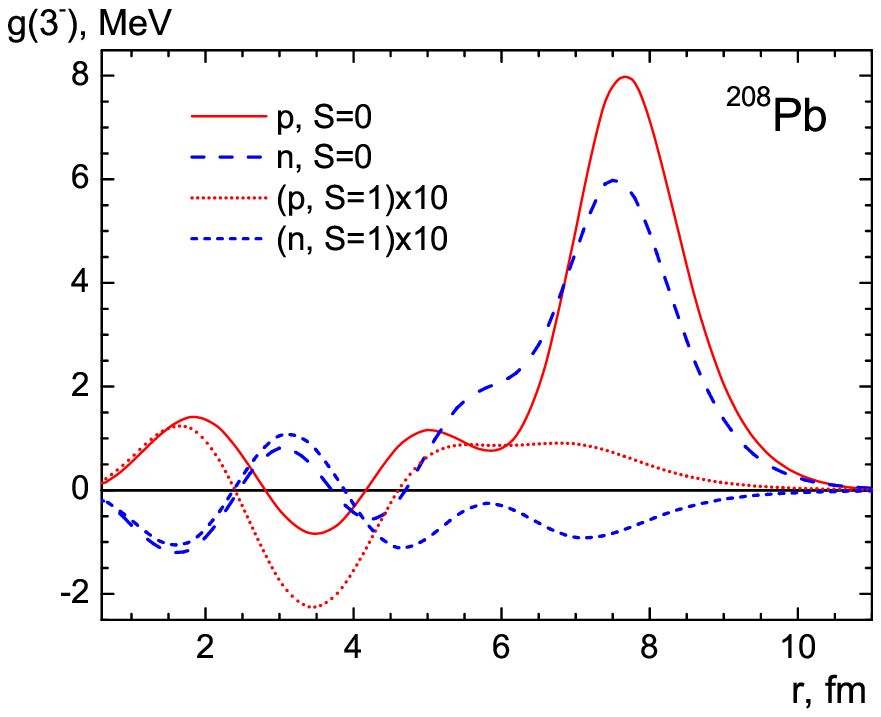}}
\vspace{2mm} \caption{ Components of the vertex $g_L$,
$L^{\pi}=3^-$, in $^{208}$Pb.}
\end{figure}

\begin{figure}[]
\vspace{10mm} \centerline {\includegraphics [width=86mm]{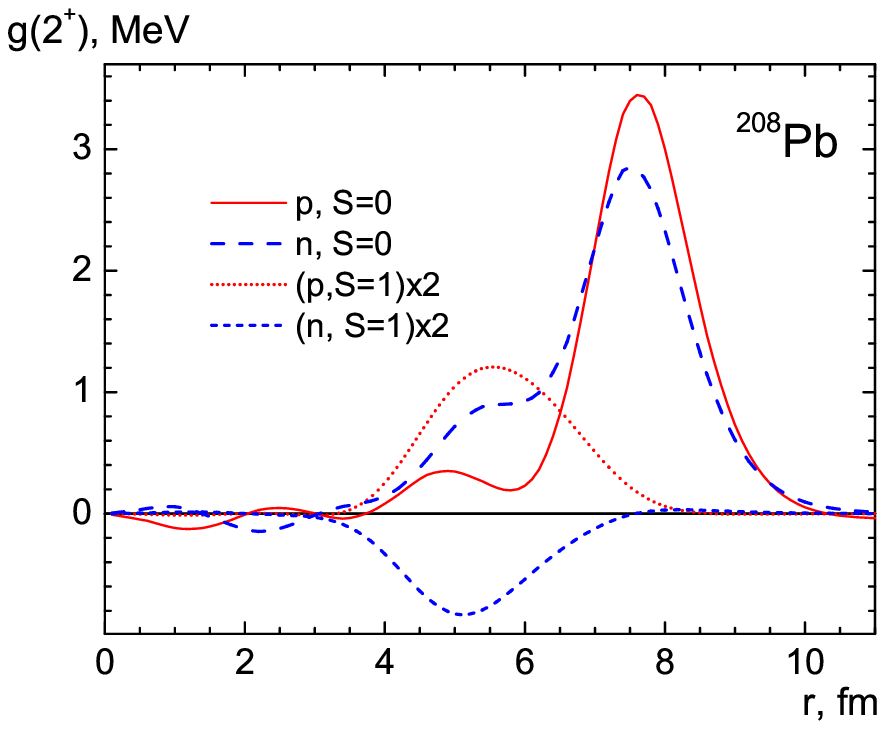}}
\vspace{2mm} \caption{ Components of the vertex $g_L$,
$L^{\pi}=5^-_1$, in $^{208}$Pb.}
\end{figure}

\begin{figure}[]
\vspace{10mm} \centerline {\includegraphics [width=86mm]{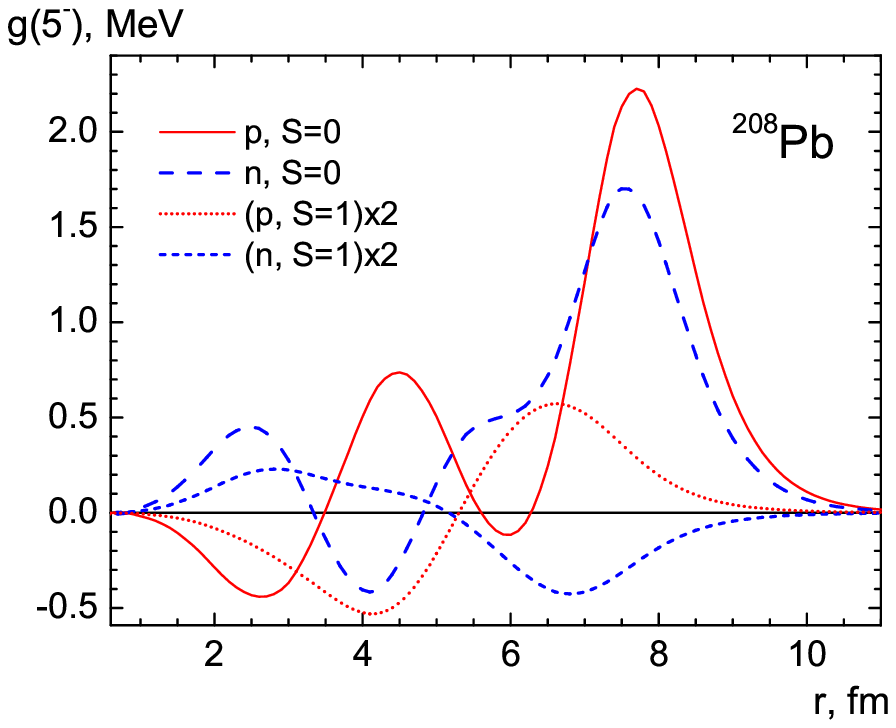}}
\vspace{2mm} \caption{ Components of the vertex $g_L$,
$L^{\pi}=2^+_1$, in $^{208}$Pb.}
\end{figure}

Let us now calculate magnetic moments and corresponding gyromagnetic
ratios $g_L^{\rm ph}$ of the $L$-phonons in $^{208}$Pb. It is given
by the diagram of Fig. 8, Eq. (\ref{GGG}).  We need the $g_L^{\rm
ph}$ values to evaluate the contribution of the phonon magnetic
moment term, Fig. 6 and Eq. (\ref{GDD2}). In addition, this
calculation elucidates to a large extent the structure of the
phonons under consideration. The results are given in Table II.  We
showed separately the $j$- and $s$-components, according to Eq.
(\ref{mu0}), with obvious substitution of ${\bf l} = {\bf j}-{\bf
s}$. The subscripts $n$, $p$ refer to the neutron and proton
subsystems whereas $L$ corresponds to their sum, e.g.
$\mu_L^{(j)}=\mu_n^{(j)}+\mu_p^{(j)}$.
   To understand better the nature of phonons we
investigate it is worth comparing the $g_L^{\rm ph}$ values obtained
with the BM model prediction $g_{L,{\rm BM}}^{\rm ph}=Z/A=0.394$. We
see that only for the $3^-$ and $4^+_2$-states our values are rather
close to the BM one whereas in other cases there is nothing in
common with these two theory predictions. Note that in the BM model
the spin-component $\mu_L^{(s)}$ is absent. If we neglect in Eq.
(\ref{GGG}) the spin term of the effective field $V(M1)$ and put
$\zeta_l=0$, i.e. take $V(M1)=j$, we obtain the BM value of
$g_L^{\rm ph}$ for all the states under consideration. Thus, the
microscopic value of the gyromagnetic ratio deviates from the
classic model prediction due to the spin term and non-zero value of
$\zeta_l$.

To testify the formulas above and estimate the accuracy of the
calculations, it is instructive to apply Eqs. (\ref{GGG}),
(\ref{GGG1}) to the spurious phonon $L^{\pi}=1^-$, $\omega_{1^-}=0$.
In this case, the term $g_{11}$ in the sum of Eq. (\ref{gLS01})
vanishes, whereas the term $g_{10}(\omega)$ is singular at
$\omega=0$ \cite{KhS}, \beq g_{10}(\omega) =\frac 1 {\sqrt{2\omega
B_1}} \frac {\partial U } {\partial r}, \label{g1}  \eeq where
$U(r)$ is the central part of the mean-field potential generated by
the energy functional (\ref{E0}) and  $B_1=3m/(4\pi A)$ is the BM
mass coefficient \cite{BM1}. Eq. (\ref{g1}) follows from the exact
TFFS self-consistency relation \cite{Fay-Khod} with some
simplifications and neglecting  the spin-orbit terms. The
singularity in all the above expressions containing $g_1^2$ is
compensated as the corresponding integrals of the Green functions
are proportional to $\omega$. This approximation for $g_1$ violates
a little the normalization relation (\ref{norm}) leading to the
value -1.074 instead of -1. If we correct normalization, we obtain
for the magnetic moment of the spurious $1^-$-phonon we obtain
$\mu(1^-)=0.403$ in very good agreement with the BM gyromagnetic
ratio 0.394. This calculation confirms the self-consistency of the
scheme developed above.

In our calculations, the spin component is negligible for the
$3^-$-state only. It confirms that this state, indeed, is most
similar to the BM surface vibrations. The phonon creation amplitudes
$g_L$ are displayed in Figs. 11-13 for the states $3^-$, $5^-_1$ and
$2^+_1$, respectively. We see that all of them contain the BM-like
($\propto {\partial U } / {\partial r}$) surface maxima of the
spin-zero components which are significantly larger than components
with $S=1$. However, for $5^-_1$ and $2^+_1$ states they are not
negligible. Moreover, they possess maxima at $r\simeq 6$ fm where
the wave functions of the single-particle states in vicinity of
Fermi level have their maxima too. This is true for couples of the
wave functions $(\phi_{\lambda}\phi_{\lambda'})$  which correspond
to the term with a small energy denominator in Eq. (\ref{GGG1})
which always exists for the phonon states under discussion.
At the same time the maxima of the main, $S=0$, components are
shifted to the right at $\simeq 1\div 2$ fm  due to the coherent
contribution of a lot of single-particle states distant from Fermi
surface. However, in this region the wave functions under discussion
began vanish. As the result, for these states the contribution of
the $S=1$ to the phonon magnetic moment is often rather big.

\section{j-conservation condition}

As the total angular momentum ${\bf j} = {\bf l}+ 1/2 {\bfg \sigma}$
is the integral of motion, the identity \beq V({\bf j}) = {\bf j}
\label{j-cons1} \eeq should be fulfilled. Within the standard TFFS,
it leads to relations for the local charges $e_q^{n,p}$ \cite{AB1}.
Obviously, when the PC corrections are included, the sum of all of
them discussed above should vanish in the case of $V_0={\bf j}$. As
it was mentioned in Sect. 3, the sum of the first four terms of
(\ref{PC-sum}) is equal to zero for any value of $L$. This can be
shown analytically and by numerical calculations as well.

This means that the identity \beq \delta V_{\rm non-pole}({\bf j}) =
- \delta V^{(2)}_{GDD}({\bf j}) \label{non-pole}\eeq should be
fulfilled to guarantee the identity (\ref{j-cons1}). We suppose that
the similar relation, \beq \delta V_{\rm non-pole}(M1) = - \delta
V^{(2)}_{GDD}(M1), \label{non-pole}\eeq is valid for the complete
$M1$-field. This relation is the main ingredient of the model we
use.

It is worth noting that the ansatz (\ref{final}) violates the ${\bf
j}$-conservation law as it takes into account higher order terms  in
$g_L^2$ in an non-consistent way. However, this violation turned out
to be rather small.

\section{Magnetic moments of the odd neighbors of $^{208}$Pb}

\begin{table}[ht]
\caption{Contributions of different low-lying phonons to the
$Z$-factors of single-particle states in $^{208}$Pb.}
\begin{tabular}{c c c c c c }

\hline \hline\noalign{\smallskip} $\lambda$  & $Z(3^-_1)$&  $Z(5^-_1)$ &$Z(2^+_1)$ &$Z'$ & $Z_{\rm tot}$\\
\noalign{\smallskip}\hline\noalign{\smallskip}

$3d_{5/2}^n $   &  0.963 &   0.989  &  0.967  &  0.944  &  0.869 \\

$1j_{15/2}^n$   &  0.657 &   0.985  &  0.989  &  0.974  &  0.623 \\

$1i_{11/2}^n$   &  0.974 &   0.998  &  0.990  &  0.981  &  0.944 \\

$2g_{9/2}^n$    &  0.918 &   0.995  &  0.989  &  0.975  &  0.881 \\

$3p_{1/2}^{-n}$ &  0.963 &   0.995  &  0.986  &  0.979  &  0.925 \\

$3p_{3/2}^{-n}$ &  0.960 &   0.992  &  0.984  &  0.971  &  0.910 \\

$2f_{5/2}^{-n}$ &  0.964 &   0.997  &  0.985  &  0.974  &  0.922 \\

$1i_{13/2}^{-n}$&  0.943 &   0.990  &  0.990  &  0.979  &  0.904 \\
\noalign{\smallskip}
$2f_{5/2}^p   $   & 0.953    &  0.995   &  0.941 &   0.907 &  0.810 \\

$1i_{13/2}^p  $   &  0.747   &  0.992   &  0.993 &   0.975 &  0.718 \\

$2f_{7/2}^p   $   &  0.898   &  0.996   &  0.991 &   0.971 &  0.861 \\

$1h_{9/2}^p   $   &  0.978   &  0.999   &  0.996 &   0.987 &  0.960 \\

$3s_{1/2}^{-p}$   &  0.959   &  0.996   &  0.992 &   0.979 &  0.928\\

$2d_{3/2}^{-p}$   &  0.964   &  0.998   & 0.992  &  0.982  &  0.937\\

$1h_{11/2}^{-p}$  &  0.951   &  0.996   & 0.995  &  0.982  &  0.926\\

$2d_{5/2}^{-p}$   &   0.758  &  0.996   & 0.986  &  0.968  &  0.721\\

\noalign{\smallskip}\hline \hline
\end{tabular}\label{tab3}
\end{table}

\begin{table*}[]
\caption{PC contributions  to the magnetic moments (in $\mu_N$) of
odd neighbors of $^{208}$Pb. The notation is explained in the text.}

\begin{tabular}{c c c c c c c c c c }

\hline \hline\noalign{\smallskip} nucleus &$\lambda$  & $\mu_0$&
$L^{\pi}$  &$\delta \mu^Z$& $\delta \mu_{GGD}$ & $\delta
\mu_{GDD}^{(1)}$& $(\delta
\mu_{GDD}^{(2)})$ & $\delta \mu_{\rm end}'$& $\delta \mu_{\rm PC} $ \\

\noalign{\smallskip}\hline\noalign{\smallskip}

$^{207}$Pb & $3p_{1/2}^{-n} $ & 0.474   &   $3^-_1$ &  -0.018&0.015
                                                         & -0.2E-3 &0.018  &0.5E-3  &-0.003   \\

&          &    &all      & -0.037  &  0.030    &   0.006      & 0.031  &0.002   & 0.001    \\
\noalign{\smallskip}\hline\noalign{\smallskip}
$^{207}$Pb & $2f_{5/2}^{-n} $   & 0.720 &   $3^-_1$& -0.027    & 0.019  & -0.001 & 0.071 & -0.002 & -0.011\\
                 &          & &all      & -0.058    & 0.046  & 0.050  & 0.140  &-0.002 &  0.036  \\
\noalign{\smallskip}\hline\noalign{\smallskip}

$^{209}$Pb & $2g_{9/2}^n $   & -1.337&  $3^-_1$ &0.120 &-0.111  & -0.051  &0.166 & 0.007& -0.035 \\
                 &              & &all          &0.174 &-0.154  & -0.013  &0.157 & 0.008&  0.015  \\
\noalign{\smallskip}\hline\noalign{\smallskip}

$^{207}$Tl & $3s_{1/2}^{-p} $& 1.857    &  $3^-_1$     &-0.082   &0.089  &-0.004  &0.005 &-0.004 &-0.001 \\
                 &          &                &all      &-0.137   &0.130  &-0.005  &0.001 & 0.035  &0.023 \\
\noalign{\smallskip}\hline\noalign{\smallskip}

$^{209}$Bi & $1h_{9/2}^p $   & 3.691    & $3^-_1$   & -0.083   &0.051  &0.014  &0.048 &0.002  &-0.015  \\
                  &          &          &all        & -0.146   &0.096  &0.033  &0.041 &0.002  &-0.015    \\
\noalign{\smallskip}\hline \hline\noalign{\smallskip}

$^{209}$Pb & $1j_{15/2}^n $   & -1.315    & $3^-_1$     & 0.688   &-0.679  &0.696  &-0.237 &-0.001  & 0.703  \\
                  &                    &   &all         & 0.754   &-0.738  &0.780  &-0.249 &-0.001  & 0.794  \\
\noalign{\smallskip}\hline\noalign{\smallskip}

$^{209}$Bi &  $1i_{13/2}^p $   & 8.071 & $3^-_1$     & -2.730   & 1.530  &0.413  &-0.162 &0.3E-3  & -0.785  \\
                  &           &          &all        & -3.020   & 1.730  &0.460  &-0.178 &0.001   & -0.824 \\

\noalign{\smallskip}\hline \hline

\end{tabular}\label{tab4}
\end{table*}

As it was mentioned above the  first two terms of Eq. (\ref{PC-sum})
dominate. Let us begin from the ``end correction'' which is simpler
for the analysis. In Table III we show contributions of three
phonons to the $Z$-factors of different single-particle states in
the odd neighbors of $^{208}$Pb, i.e.  in $^{207,209}$Pb isotopes
for neutrons and $^{207}$Tl, $^{209}$Bi nuclei for protons. Here, we
used the notation $Z_{\rm tot}=\prod_{i=1}^9 Z_i$ and
$Z'=Z_3\prod_{i=5}^9 Z_i$, where the index $i$ numerates all the
collective states listed in Table I.  Thus, the states $3^-_1$,
$5^-_1$ and $2^+_1$ are excluded in $Z'$.

\begin{table}[ht]
\caption{Magnetic moments (in $\mu_N$ units) of the odd neighbors of
$^{208}$Pb.}
\begin{tabular}{c c c c c c}

\hline \hline\noalign{\smallskip} nucleus &$\lambda$  & $\mu_0$&
${\delta \mu}$& $\mu_{\rm th}$ &$\mu_{\rm exp}$\\
\noalign{\smallskip}\hline\noalign{\smallskip}

$^{207}$Pb \;&\; $3p_{1/2}^{-n} $  \;\;  &\;\; 0.474  \;\;&\;\;0.001\;\;&\;\; 0.475 \;\;& 0.592585(9) \\

$^{207}$Pb &  $2f_{5/2}^{-n} $    & 0.720  &0.036    & 0.756  & 0.80(3) \\

$^{209}$Pb &  $2g_{9/2}^n $       &-1.337  &0.015    &-1.322  & -1.4735(16) \\

$^{209}$Pb & $1j_{15/2}^n $   & -1.315    & 0.794    &-0.521  &   -    \\

$^{207}$Tl & $3s_{1/2}^{-p} $    & 1.857  &0.023    & 1.880  & 1.876(5) \\

$^{209}$Bi & $1h_{9/2}^p $       & 3.691  &-0.015   & 3.676  & 4.1106(2) \\

$^{209}$Bi &  $1i_{13/2}^p $     & 8.071 & -0.824   & 7.247  &    -   \\
\noalign{\smallskip}\hline \hline
\end{tabular}\label{tab5}
\end{table}

As it was stated in the Introduction, the contribution of the $3^-$
state to the PC corrections dominates in magic nuclei due to its
strong collectivity. Table III confirms this statement on average,
although often the the sum of the contributions of all other phonons
to the difference of $(1-Z_{\lambda})$ is comparable with the one of
the $3^-$ phonon alone. From the other phonons, the  $2^+_1$ phonon
contributes most to the $Z$-factor  although it is less collective
than $5^-_1, 5^-_2$ phonons. The reason becomes clear if one looks
to Eq. (\ref{ddSig}) which determines the phonon contribution to the
difference of $(1-Z_{\lambda})$, see also Eq. (\ref{betL}) which
gives an estimate of a separate contribution to different PC
correction. It is obvious that the situation is on average more
preferable for phonons of positive parity. Indeed, in this case the
states $\lambda, \lambda'$ in (\ref{betL}) are of the same parity
and it is easier to find those single-particle states with a ``small
denominator''. That is why there are two cases, the $3d_{5/2}^n$ and
$2f_{5/2}^p$ states, where the contribution of the $2^+_1$ phonon to
the difference of $(1-Z_{\lambda})$ is comparable with the one of
the $3^-$ phonon, although the matrix elements of $g_2$ are
typically   three times smaller than those of $g_3$. Typical value
of the $Z$-factor in $^{208}$Pb is $Z_{\rm tot}=0.8\div 0.9$. In
this case, the perturbation theory in $g^2$ should work well.
However, there are one neutron state, $1j^n_{15/2}$, and two proton
ones, $1j^p_{13/2}$ and $2d^{-p}_{5/2}$, for which we have $Z_{\rm
tot}=0.7 \div 0.6$ and application of the perturbation theory is
questionable.

Table IV contains various PC corrections to magnetic moments of odd
neighbors of $^{208}$Pb. In agreement with the above ansatz, the
total value of $\delta \mu$ given in the last column is the sum of
the terms in the first line of Eq. (\ref{PC-sum}). The term $\delta
\mu_{GDD}^{(2)}$ is presented just for information. In the upper
part of the table five cases are presented for which experimental
data exist. We see that for all of them the PC correction to $\mu$
value is negligible. The reason is the strong cancelation of the
first two corrections, $\delta \mu^Z$ and $\delta \mu_{GGD}$. The
term $\delta  \mu_{GDD}^{(1)}$ is usually significantly smaller than
these two main ones.  The term $\delta \mu_{\rm end}'$ is defined in
accordance with Eq. (\ref{end3}). This correction is always
negligible and is included just for completeness.

In the two bottom lines of Table IV we presented two examples of
strong PC corrections for two states which magnetic moments are not
known. In both  cases, the PC correction to the $Z$-factor is rather
big, see Table III, because of the existence of a term with small
energy denominator, see Eq. (\ref{betL}), for the $3^-$-phonon. Due
to the same reason, the term $\delta  \mu_{GDD}^{(1)}$ is also
rather big. In the case of the $1j_{15/2}^n $ state, $\delta \mu^Z$
and $\delta \mu_{GGD}$ terms again cancel each other, but the
$\delta \mu_{GDD}^{(1)}$ term  is comparable with the first two
terms and leads to a big total correction. In the $1i_{13/2}^p $
case, the cancelation is not so strong as usual, but  the $\delta
\mu_{GDD}^{(1)}$ term is again important.

Table V contains the final results for magnetic moments, \beq
\mu_{\rm th}=\mu_0+\delta \mu, \label{mu_theor} \eeq  where the
first term is the TFFS prediction for the magnetic moment, i.e. the
solution of Eq. (\ref{Vef}), and $\delta \mu$ is the sum of all the
PC corrections considered above. We see that they can not help in
solving the long-standing problem of the theoretical description of
the ground state magnetic moment of the $^{209}$Bi nucleus.
Evidently, in this case the usual TFFS form (\ref{Fs}) for the
spin-dependent LM amplitudes is not sufficient and the spin-orbit
terms and tensor ones in cross channel should be included. This is
rather difficult in the coordinate space method we use for solving
Eq. (\ref{Vef}). The direct solution in the $\lambda$-representation
in a large basis would be more adequate in this case. Importance of
the spin-orbit force for this problem was found in Ref. \cite{DmT}.

In conclusion of this Section, we repeat that our model for PC
contributions to  magnetic moments of odd neighbors of $^{208}$Pb
leads to very small corrections. A similar result was obtained
previously by Tselyaev \cite{Tsel} who considered mainly diagrams we
omit, displayed in Fig. 9 and Fig. 10. The reason was in strong
cancelation of these diagrams similar to cancelation of the diagrams
in Fig. 3 and Fig. 5 within our model.

\section{Magnetic moments of semi-magic nuclei}

Let us go to semi-magic nuclei. As it was mentioned in the
Introduction, we deal with such odd nuclei where the odd nucleon
belongs to the non-superfluid subsystem. Otherwise, all equations of
Sect. 3 should be generalized by including the superfluidity. In
particular, we consider the odd-proton neighbors of the even lead
nuclei, i.e. Tl and Bi isotopes, and the same for the tin-core
region, where we analyze In and Sb isotopes. However,  the neutron
sub-system which is now superfluid and we have to apply the vector
notation of Eq. (\ref{gL_s}) using the initial notation $g_L^{(0)}$
for the normal component of $\hat g_L$ and $g_L^{(1,2)}$ for the
anomalous ones. All the equations of Section 3 remain valid with the
only exception of Eqs. (\ref{GGG}), (\ref{GGG1}) corresponding to
Fig. 8, for the phonon magnetic moment or the gyromagnetic ration
$g_L^{\rm ph}$. Indeed, now the neutron triangle should be included
also, with taking into account the superfluidity effects. Therefore
we use the BM model prediction for $g_L^{\rm ph}$ in this Section.
This is justified as far as we consider only such phonons which are
very collective.

In all even-even spherical non-magic nuclei there exists a very
collective low-lying $2^+_1$-state with an excitation energy
$\omega_{2^+}\simeq 1$ MeV.  In fact, it is the only state which
should be taken into account for the PC corrections to magnetic
moments. For the odd-proton neighbors of nuclei $^{200-206}$Pb we
included also the $3^-_1$-state, the next in the range of
collectivity, and found that it gives only several \% of the main PC
correction of the $2^+_1$-state. Therefore in other cases  we limit
ourselves with this PC correction only. In Table IV we present
excitation energies and $B(E2)$ values of the even Pb isotopes.

\begin{table}[]
\caption{Characteristics of the low-lying $2^+_1$-phonons in even Pb
isotopes,
$\omega_2$ (MeV) and $B(E2,{\rm up)} \times 10^4 $(${\rm e^2 fm}^{4}$).}
\begin{tabular}{c c c c c  }

\hline \hline\noalign{\smallskip} $A$  & $\omega_2^{\rm th}$   &
$\omega_2^{\rm exp}$
&  $B(E2)^{\rm th}$ &  $B(E2)^{\rm exp}$  \\
\noalign{\smallskip}\hline\noalign{\smallskip}
188    & 1.028      & 0.723    & 0.551    &  -    \\
190    & 0.930      & 0.733    & 0.617    &  -    \\
192    & 0.849      & 0.853    & 0.634    &  -    \\
194    & 0.792      & 0.965    & 0.646    &  -    \\
196    & 0.764      & 1.049    & 0.627    &  -    \\
198    & 0.762      & 1.063    & 0.624    &  -    \\
200    & 0.789      & 1.026    & 0.479    &  -    \\
202    & 0.823      & 0.960    & 0.373    &  -    \\
204    & 0.882      & 0.899    & 0.250    & 0.162 (0.004)     \\
206    & 0.945      & 0.803    & 0.130    & 0.100 (0.002)     \\
208    & 4.747      &  4.086   & 0.189    & 0.30  (0.03)     \\
210    & 1.346      &  0.799   & 0.036    & 0.051 (0.015)     \\
212    & 1.443      &  0.804   & 0.131    & -     \\
214    & 1.125      &  0.836   & 0.161    & -     \\
\noalign{\smallskip}\hline \hline

\end{tabular}\label{tab6}
\end{table}

\begin{table}[]
\caption{$Z$-factor values of two proton levels close to the Fermi surface due to the $2^+_1$
and $3^-_1$-phonons in the lead chain.}
\begin{tabular}{c c c c c  }

\hline \hline\noalign{\smallskip} $A$  & $Z(3s_{1/2})[2^+]$   &
$Z(3s_{1/2})[3^-]$
&  $Z(1h_{9/2})[2^+]$  &  $Z(1h_{9/2})[3^-]$   \\
\noalign{\smallskip}\hline\noalign{\smallskip}
188    & 0.830      &  -    & 0.659    &  -    \\
190    & 0.822      &  -    & 0.614    &  -    \\
192    & 0.834      &  -    & 0.605    &  -    \\
194    & 0.719      &  -    & 0.542    &  -    \\
196    & 0.824      &  -    & 0.547    &  -    \\
198    & 0.810      &  -    & 0.526    &  -    \\
200    & 0.755      &0.966  & 0.619    & 0.980    \\
202    & 0.743      &0.967  & 0.628    & 0.981    \\
204    & 0.856      &0.968  & 0.786    & 0.982    \\
206    & 0.915      &0.965  & 0.880    & 0.981    \\
\noalign{\smallskip}\hline \hline

\end{tabular}\label{tab7}
\end{table}

\begin{figure}[]
\vspace{10mm} \centerline {\includegraphics [width=86mm]{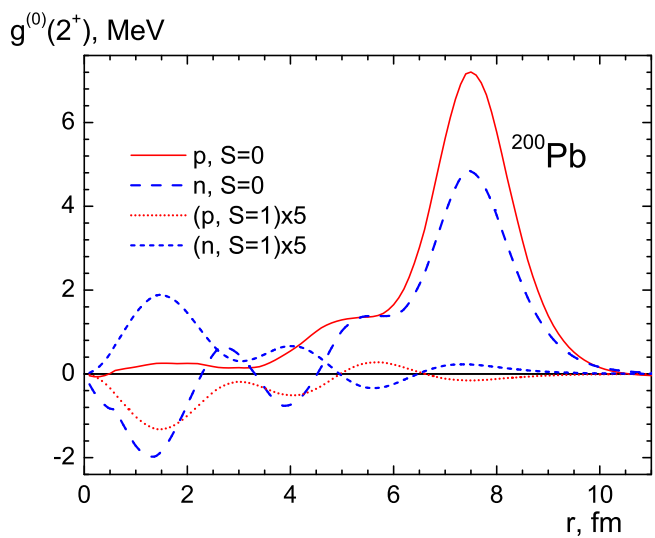}}
\vspace{2mm} \caption{ Components of the normal amplitude
$g_2^{(0)}$ in $^{200}$Pb.}
\end{figure}

\begin{figure}[]
\vspace{10mm} \centerline {\includegraphics [width=86mm]{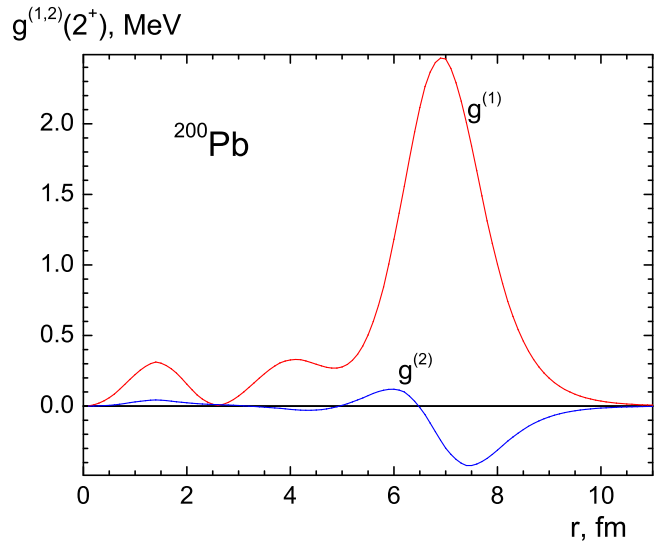}}
\vspace{2mm} \caption{ Components of the neutron anomalous
amplitudes $g_2^{(1)}$ and $g_2^{(2)}$ in $^{200}$Pb.}
\end{figure}

These results were reported previously in \cite{BE2}. Experimental
data on $B(E2)$ values are not complete whereas the $\omega_{2^+}$
value is known  for all isotopes in the chain we consider. As the
analysis in the previous Section showed the latter is of primary
importance for calculating the PC corrections. We see that the
excitation energies on average are in reasonable agreement with the
data. However, the theory does not reproduce the experimental trend.
Specifically, if we exclude the magic $^{208}$Pb nucleus, the
experimental $\omega_{2^+}$ values have a maximum, $\simeq 1\;$MeV,
in the middle of the chain, $A\simeq 200$, and decrease  to $\simeq
0.7\div 0.8\;$MeV at both  ends. The theoretical predictions  have
opposite behavior. They show in the middle a minimum, $\simeq
0.8\;$MeV, and increase up to $\simeq 1\;$MeV in the left end and up
to $\simeq 1.4\;$MeV in the right one. Therefore we may expect that
our predictions for $\delta \mu_{\rm PC} $ values will be too high
in the middle of the chain and too low in the both ends.   However,
as the estimation (\ref{betL}) shows, the position of
single-particle levels in the Fermi surface vicinity are often as
important as the $\omega_L$ value. In the very right end of the
chain the difference of $\omega_2^{\rm th} - \omega_2^{\rm exp}$
reaches so large value as 0.6 MeV. In view of such deficiency, we
may wait most errors of our predictions for PC corrections in this
part of the chain.

In Fig. 17 we display the components of the normal amplitude
$g_L^{(0)}$ in $^{200}$Pb which is in the middle of the lead chain.
For graphical convenience the spin-one components are multiplied by
the factor of 5. Remind that, in fact, we deal with the proton
vertices only. We see that, just as for the $3^-$-state in
$^{208}$Pb, the spin-one component is negligible and can be
neglected. For completeness we present also in Fig. 18 the neutron
anomalous amplitudes $g_L^{(1)}$ $g_L^{(2)}$, spin-zero components
only. We see that the first of them is rather big, comparable to the
normal vertex $g_L^{(0)}$. Note that all the components of the
vector $\hat g_L$ are normalized with the same normalization
condition (\ref{norm}) and can be compared directly.

To estimate the scale of the PC corrections, it is instructive to
analyze the $Z$-factors of the states we consider. We deal with the
odd Tl isotopes with $3s_{1/2}$ state for the odd proton and the odd
Bi isotopes with $1g_{9/2}$ state for the odd proton. Corresponding
values of the $Z$-factors induced by the $2^+_1$-state are given in
Table VII. For lead cores $A=200\div 206$ we showed also  the
$Z$-factor induced by the $3^-_1$-state. We see that it is rather
close to unit, so that $Z_{\rm tot}=Z(2^+)Z(3^-)\simeq Z(2^+)$. This
is a signal that all the PC corrections due to the $3^-_1$-state are
now very small and those from the $2^+_1$-state could be taken into
account only.

We see that for the major part of the lead isotopes the $Z$-factors
deviate from unit significantly, $Z\simeq 0.5\div 06$. This occurs
because in the sum of Eq. (\ref{ddSig}) which determines the
$Z$-factor there is one or two ``dangerous'' terms with small energy
denominators for which the perturbation parameter (\ref{betL}) is
not small. In such a situation, the perturbational in $g^2_L$ theory
becomes doubtful and one has to consider higher order corrections.
There is a simple way to take into account some of them. Let us
separate the PC correction to the effective field, the first line of
Eq. (\ref{PC-sum}), into two components, \beq \delta V= \delta V^Z +
\delta V', \eeq and the PC correction to the magnetic moment, \beq
\delta \mu = \delta \mu^Z + \delta \mu'. \eeq Instead of the direct
perturbation formula for the magnetic moment (\ref{mu_theor}) we
will use also the following ansatz: \beq \tilde \mu_{\rm th}=
Z_{\lambda}(\mu_0+\delta \mu'). \label{mu-tild} \eeq The physical
meaning of this recipe is very simple. It corresponds to introducing
the ``end correction'' similar to that of Fig. 3 not only to the
effective field $V(M1)$ itself, but also to triangle diagrams of
Fig. 5 and Fig. 6 describing other types of the PC corrections. Of
course, this is just an intuitive ansatz and a consistent way to
consider the dangerous terms is necessary. Up to now it is not
developed and we present in Fig. 19 both  predictions for Tl
isotopes  of Eq. (\ref{mu_theor}) corresponding to the
$g_L^2$-approximation and of the ansatz (\ref{mu-tild}) where also
some higher in $g_L^2$ terms are considered. In this case, the
absolute value of the PC correction of Eq. (\ref{mu-tild}) is
reduced, although  the difference between these two results is
rather small.

We see that for all isotopes of this chain except $^{207}$Tl, i.e.
all non-magic ones, the sign of the PC correction is negative which
always helps to reproduce the data. The experimental $\mu$ value
undergoes a sharp jump from $^{207}$Tl to $^{205}$Tl remaining
almost constant for lighter isotopes. The PC correction describes
such a behavior only qualitatively. Its absolute value grows
smoothly till $^{199,201}$Tl and becomes again rather small for
lighter isotopes. Table VIII shows separate components of $\delta
\mu$, with the same notation as in Table V. Again, as it took place
for the odd neighbors of the magic $^{208}$Pb, the resulting $\delta
\mu$ value is much smaller than the terms $\delta \mu^Z$ and $\delta
\mu_{GGD}$ separately which have opposite signs and compensate each
other strongly. Other components of $\delta \mu$ again are less
important. However, the sign of the $\delta  \mu_{GDD}^{(1)}$ term
is the same as of $\delta \mu_{GGD}$ and ``helps'' to diminish the
total $\delta \mu$ value.

Similar results for Bi isotopes are displayed in Fig. 20 and
separate terms of $\delta \mu$ are presented in Table IX. Now the
cancelation of different  PC corrections is even  stronger than for
the Tl isotopes, and the $|\delta \mu|$ value is usually 10 or 15
times less than $|\delta \mu^Z|$. Contrary to the Tl case, now the
sign of the PC correction looks incorrect. However, the main reason
for the contradiction between the theory and experiment is evidently
incorrect value of $\mu_0$. Indeed, in $^{209}$Bi where the PC
correction is negligible, see Section 5, we have $\mu_{\rm
exp}-\mu_0\simeq 0.5\; \mu_N$. As it was discussed above, the main
reason for this deviation is supposedly absence of the spin-orbit
and additional tensor terms in the spin-dependent LM interaction
(\ref{Fs}). In any case, the final conclusion about the sign of the
PC corrections could be made only after solving the problem of
$^{209}$Bi.

\begin{figure}[]
\vspace{10mm} \centerline {\includegraphics [width=86mm]{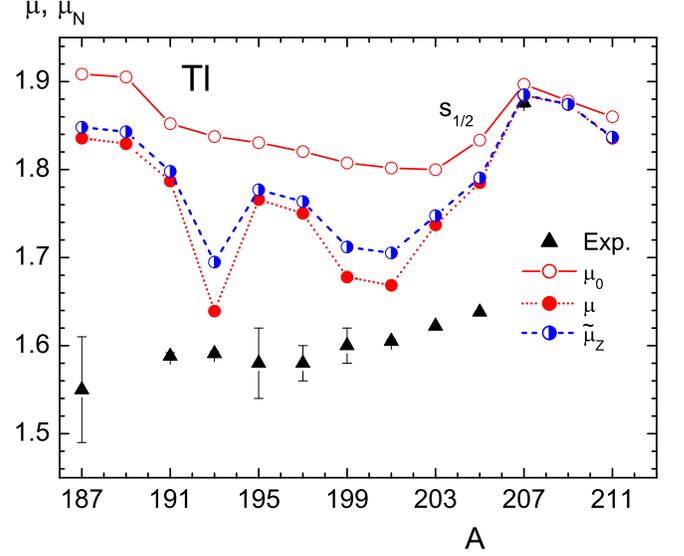}}
\vspace{2mm} \caption{Magnetic moments of Tl isotopes.}
\end{figure}

\begin{figure}[]
\vspace{10mm} \centerline {\includegraphics [width=86mm]{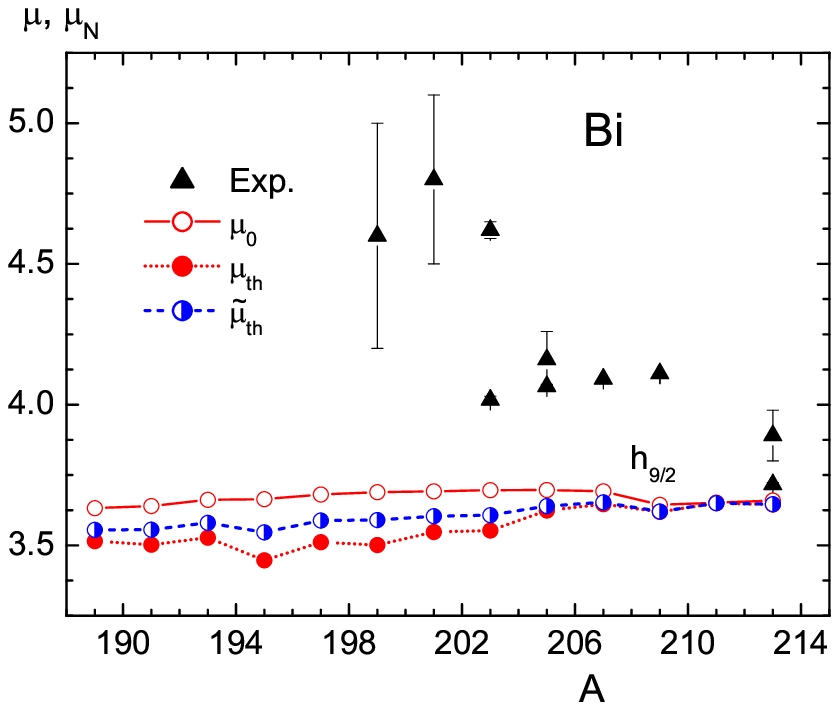}}
\vspace{2mm} \caption{(Magnetic moments of Bi isotopes.}
\end{figure}

\begin{table*}[]
\caption{PC contributions  to the magnetic moments (in $\mu_N$) of
odd Tl isotopes, $\lambda_0=3s_{1/2}^{-p}$. The notation is
explained in the text nearby Table IV.}

\begin{tabular}{c c c c c c c c c c c c}

\hline \hline\noalign{\smallskip} $A$ &$\mu_{\rm exp}$ & $\mu_0$&
$L^{\pi}$  &$\delta \mu^Z$& $\delta \mu_{GGD}$ & $\delta
\mu_{GDD}^{(1)}$& $(\delta
\mu_{GDD}^{(2)})$ & $\delta \mu_{\rm end}'$& $\delta \mu_{\rm tot} $ &$\mu_{\rm th}$ &$\tilde \mu_{\rm th}$\\

\noalign{\smallskip}\hline\noalign{\smallskip}

  207 & +1.876(5)     &1.857 &all       &-0.137 &0.130  &-0.005 &0.001 & 0.035  &0.023 &1.884 &1.885\\

  205 &+1.63821461(12)&1.834 &$2^++3^-$ &-0.236 &0.175  &0.014  &0.022 &-0.003  &-0.048&1.785 &1.791 \\

  203 &+1.62225787(12)&1.800 &$2^++3^-$ &-0.363 &0.281  &0.022  &0.027 &-0.003  &-0.063&1.737 &1.748 \\

  201 &+1.605(2)      & 1.802&$2^++3^-$ &-0.684 &0.505  &0.049 &0.057  &-0.001  &-0.133 &1.669&1.705 \\

  199 &+1.60(2)       & 1.808&$2^++3^-$ &-0.649 &0.475  &0.047 &0.057  &-0.003  &-0.130 &1.678&1.712 \\

  197 &+1.58(2)       & 1.820&$2^+$     &-0.426 &0.332  &0.025 &0.036  & 0.0    &-0.070 &1.750&1.764 \\

  195 &+1.58(4)       & 1.831&$2^+$     &-0.390 &0.303  &0.023 &0.034  & 0.0    &-0.065 &1.766&1.777 \\

  193 &+1.5912(22)    & 1.838&$2^+$     &-0.717 &0.442  &0.076 &0.089  & 0.001  &-0.198 &1.639 &1.695 \\

  191 & 1.588(4)      & 1.852&$2^+$     &-0.370 &0.281  &0.023 &0.032  & 0.0    &-0.065 &1.787 &1.798 \\

  189 &  -            & 1.905&$2^+$     &-0.413 &0.310  &0.027 &0.034  & 0.0    &-0.076 &1.829 &1.843 \\

  187 &  -            & 1.908&$2^+$     &-0.391 &0.292  &0.026 &0.031  & 0.0    &-0.073 &1.836 &1.848 \\

\noalign{\smallskip}\hline \hline

\end{tabular}\label{tab8}
\end{table*}

\begin{table*}[]
\caption{PC contributions  to the magnetic moments (in $\mu_N$) of
odd Bi isotopes, $\lambda_0=1g_{9/2}^p$. The notation is explained
in the text nearby Table IV.}

\begin{tabular}{c c c c c c c c c c c c}

\hline \hline\noalign{\smallskip} $A$ &$\mu_{\rm exp}$ & $\mu_0$&
$L^{\pi}$  &$\delta \mu^Z$& $\delta \mu_{GGD}$ & $\delta
\mu_{GDD}^{(1)}$& $(\delta
\mu_{GDD}^{(2)})$ & $\delta \mu_{\rm end}'$& $\delta \mu_{\rm tot} $ &$\mu_{\rm th}$ &$\tilde \mu_{\rm th}$\\

\noalign{\smallskip}\hline\noalign{\smallskip}

213 & +3.717(13)   &3.659     & $2^+$ & -0.196 &0.170 &0.013 &-0.013 &-0.001 &-0.014&3.645 &3.646 \\

211 & 3.5(3)       &3.651     & $2^+$ & -0.029 &0.026 &0.002 &-0.001 &0.0 &-0.002 &3.650 &3.650   \\

209 & +4.1106(2)   & 3.691    &all    & -0.146 &0.096  &0.033  &0.041 &0.002  &-0.015 &3.619  &3.620 \\

207 & +4.0915(9)&3.692 &$2^++3^-$ &-0.575 &0.485 &0.043 & 0.001  &0.002 &-0.045 &3.647 &3.653  \\

205 & +4.065(7) &3.696 &$2^++3^-$ &-1.072 &0.922 &0.074 &-0.032  &0.003 &-0.073 &3.624 &3.640  \\

203 & +4.017(13)&3.696 &$2^++3^-$ &-2.263 &1.965 &0.151  &-0.109 &0.004  &-0.143 &3.553 &3.607  \\

201 & 4.8(3)    &3.692 &$2^++3^-$ &-2.344 &2.037 &0.158 &-0.109 &0.004 &-0.145 &3.547 &3.604  \\

199 & +4.6(4)   &3.689 &$2^+$     &-3.330 &2.925 &0.215 &-0.218 &0.002 &-0.188 &3.501 &3.590  \\

\noalign{\smallskip}\hline \hline

\end{tabular}\label{tab9}
\end{table*}

Let us go to the tin region. Again we consider the proton-odd
neighbors of even Sn isotopes, the odd In and Sb isotopes, with
non-superfluid proton sub-systems. The calculation scheme is the
same as for the Tl and Bi chains, in particular, we use the BM
values for the phonon gyromagnetic ratio. On the base of experience
of calculations in the lead region, in all superfluid tin nuclei  we
consider the $2^+_1$ states only except for the magic $^{100}$Sn and
$^{132}$Sn where we include also the $3^-_1$ state. In Table X the
characteristics of the $2^+_1$ states, excitation energies and
$B(E2)$ values, are given. They, just as those for the lead
isotopes, were reported  in \cite{BE2}. As it is follows from the
above analysis, the excitation energy  $\omega_2$ is of primary
importance for accuracy of calculations of the PC corrections. We
see that, with several exceptions, $\omega_2$ value is reproduced
with accuracy $100\div 200\;$keV i.e. sufficiently well. The $B(E2)$
values characterize  the ``strength'' of the phonon coupling
amplitudes $g_L$, and they are also reproduced reasonably.

\begin{table}[]
\caption{Characteristics of the low-lying $2^+_1$-phonons in even Sn
isotopes, $\omega_2$ (MeV) and $B(E2,{\rm up)} \times 10^4 $(${\rm
e^2 fm}^{4}$).}
\begin{tabular}{c c c c c  }

\hline \hline\noalign{\smallskip} $A$  & $\omega_2^{\rm th}$   &
$\omega_2^{\rm exp}$
&  $B(E2)^{\rm th}$ &  $B(E2)^{\rm exp}$  \\
\noalign{\smallskip}\hline\noalign{\smallskip}
102    & 1.453      & 1.472    & 0.065    &  -    \\
104    & 1.388      & 1.260    & 0.107    &  -    \\
106    & 1.316      & 1.207    & 0.142    & 0.195 (0.039)     \\
108    & 1.231      & 1.206    & 0.155    & 0.222 (0.019)     \\
110    & 1.162      & 1.212    & 0.188    & 0.220 (0.022)     \\
112    & 1.130      & 1.257    & 0.197    & 0.240 (0.014)     \\
114    & 1.156      & 1.300    & 0.193    & 0.24 (0.05)     \\
116    & 1.186      & 1.294    & 0.182    & 0.209 (0.006)     \\
118    & 1.217      & 1.230    & 0.172    & 0.209 (0.008)     \\
120    & 1.240      & 1.171    & 0.152    & 0.202 (0.004)     \\
122    & 1.290      & 1.141    & 0.158    & 0.192 (0.004)     \\
124    & 1.350      & 1.132    & 0.147    & 0.166 (0.004)     \\
126    & 1.405      & 1.141    & 0.120    & 0.10 (0.03)     \\
128    & 1.485      & 1.169    & 0.094    & 0.073 (0.006)     \\
130    & 1.610      & 1.221    & 0.055    & 0.023 (0.005)     \\
132    & 4.327      & 4.041    & 0.104    & 0.11 (0.03)     \\
134    & 1.142      & 0.725    & 0.033    & 0.029 (0.005)     \\

\noalign{\smallskip}\hline \hline

\end{tabular}\label{tab10}
\end{table}

\begin{table}[]
\caption{$Z$-factor values of proton levels close to the Fermi
surface induced by  the low-lying phonons in the tin chain.}
\begin{tabular}{c c c c   }

\hline \hline\noalign{\smallskip} $A$  & $Z(1g_{9/2})$   & $Z(2d_{5/2})$ &  $Z(1g_{7/2})$     \\
\noalign{\smallskip}\hline\noalign{\smallskip}

100    & 0.939    & 0.910     & 0.933   \\
102    & 0.818    & 0.783     & 0.524   \\
104    & 0.716    & 0.673     & 0.344   \\
106    & 0.634    & 0.592     & 0.238   \\
108    & 0.584    & 0.549     & 0.158   \\
110    & 0.509    & 0.481     & 0.103   \\
112    & 0.482    & 0.464     & 0.458   \\
114    & 0.510    & 0.484     & 0.238   \\
116    & 0.609    & 0.582     & 0.414   \\
118    & 0.578    & 0.547     & 0.454   \\
120    & 0.614    & 0.591     & 0.607   \\
122    & 0.634    & 0.605     & 0.591   \\
124    & 0.674    & 0.648     & 0.651   \\
126    & 0.734    & 0.712     & 0.725   \\
128    & 0.800    & 0.780     & 0.798   \\
130    & 0.914    & 0.878     & 0.836   \\
132    & 0.962    & 0.923     & 0.968   \\

\noalign{\smallskip}\hline \hline

\end{tabular}\label{tab11}
\end{table}

In Table XI, the $Z$-factor values are shown for three states
$\lambda_0$ nearby the Fermi surface, just the same which are
involved in In and Sb nuclei we analyze. We see that there are
several cases where the $Z$-factors are very small, $1-Z\ll 1$. This
is a signal that there is a catastrophic term with a small
denominator in Eq. (\ref{betL}). E.g., in the $^{110}$Sn case, for
$Z(1g_{7/2})$ we have
$\eps_{1g_{7/2}}-\eps_{2d_{5/2}}-\omega_L=0.08\;$MeV. It's clear
that in such a situation the perturbation expansion in $g_L^2$, in
fact, in $\beta_L$ of Eq. (\ref{betL}) is absolutely non-realistic,
and an exact account of the dangerous terms is necessary.
Fortunately, the nuclei for which experimental data exist are out of
this catastrophic region.

Just as for the lead chain above,  we display  in Fig. 21 the
components of the normal amplitude $g_L^{(0)}$ in the $^{118}$Sn
nuclei which is chosen as a representative of the tin chain. For
graphical convenience the spin-one components are multiplied by the
factor of 5. As before, we deal with the proton vertices only. We
see that again this mode is a typical surface vibration, the
spin-one component is very small and can be neglected. For
completeness we present again in Fig. 22 the neutron spin-zero
components of the anomalous amplitudes $g_L^{(1)}$ $g_L^{(2)}$. We
see that both of them are rather large and have the surface maxima
of opposite sign. In the result, the amplitude $g_L^{\rm
an,+}=g_L^{(1)} + g_L^{(2)}$ is very small whereas the amplitude
$g_L^{\rm an,-}=g_L^{(1)} - g_L^{(2)}$ is comparable with the normal
amplitude $g_L^{(0)}$. This point is out of the scope of this
article but it worth to note that the latter combination determines
the collective phenomena in superfluid sub-system \cite{Bel1,AB1}.

\begin{figure}[]
\vspace{10mm} \centerline {\includegraphics [width=86mm]{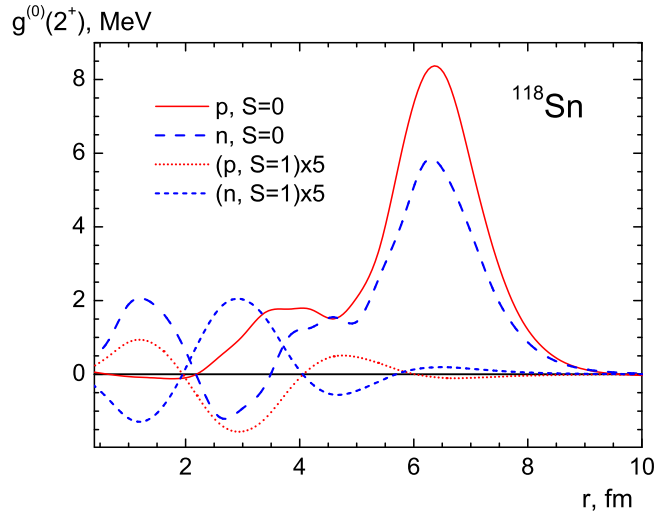}}
\vspace{2mm} \caption{Components of the normal amplitude $g_2^{(0)}$
in $^{118}$Sn.}
\end{figure}

\begin{figure}[]
\vspace{10mm} \centerline {\includegraphics [width=86mm]{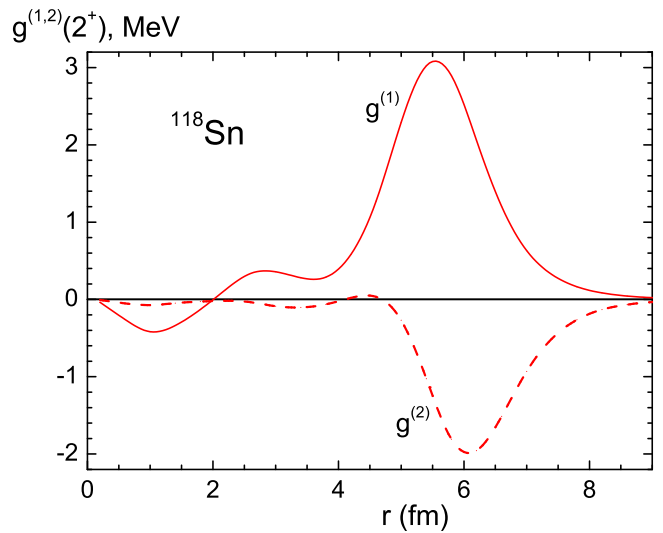}}
\vspace{2mm} \caption{Components of the neutron anomalous amplitudes
$g_2^{(1)}$ and $g_2^{(2)}$ in $^{118}$Sn.}
\end{figure}

\begin{figure}[]
\vspace{10mm} \centerline {\includegraphics [width=86mm]{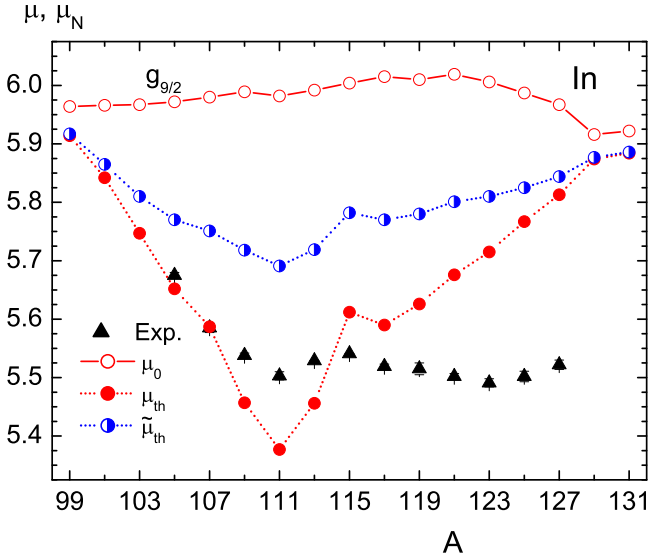}}
\vspace{2mm} \caption{Magnetic moments of In isotopes.}
\end{figure}

Let us go to PC corrections to magnetic moments and begin with odd
In nuclei. In this chain, there are 13 isotopes with known
experimental values of magnetic moments of the ground state $9/2^+$
corresponding to the $g_{9/2}^{-p}$ hole in the corresponding even
core of Sn. The theoretical predictions for the magnetic moments are
displayed in Fig. 23 whereas separate components of the PC
correction are given in Table XII. We see that the TFFS prediction
without PC corrections overestimates the experimental value by
$\simeq 0.4\; \mu_N$. The value of $\delta \mu_{\rm tot}$, column 10
of Table XII, is always negative and in general large enough to
compensate this excess. In $^{109-113}$In isotopes the PC correction
is even too large. This correlates with small values of the $Z\simeq
0.5$ which corresponds to $\partial \Sigma / \partial \eps \simeq
-1$. It is worth noting that even the biggest total PC correction
$\delta \mu_{\rm tot} \simeq - 0.6\; \mu_N$ is the result of an
almost exact compensation of the two ten times bigger corrections
$\delta \mu^Z$ and $\delta \mu_{GGD}$. Therefore any additional term
can change the results significantly. The PC correction is reduced
in this case by almost a factor two if we use the ansatz
(\ref{mu-tild}), $\mu_{\rm th}\to \tilde \mu_{\rm th}$. Remind that
it is an approximate way to take into account some higher order
terms in $g_L^2$ corresponding to partial summation of the pole end
diagrams. As it was discussed above, so large PC effect in the
$Z$-factor is a signal of a dangerous term with a small energy
denominator in the sum of Eq. (\ref{dSig2}) which determines the
$\delta \mu^Z$  and in its counterpart in the sum of Eq.
(\ref{GGD3}) for the $\delta \mu_{GGD}$ term.    More exact account
for these dangerous contributions is necessary of course. But now we
can hope that in so anomalous cases the truth is somewhere between
$\mu_{\rm th}$ and $\tilde \mu_{\rm th}$.

\begin{table*}[]
\caption{PC contributions  to the magnetic moments (in $\mu_N$) of
odd In isotopes, $\lambda_0=1g_{9/2}$, $L^{\pi}=2^+$. The notation
is explained in the text nearby Table IV.}

\begin{tabular}{  c c c c c   c c c c c  c c}

\hline \hline\noalign{\smallskip} $A$ & $\mu_{\rm exp}$ & $\mu_0$
&$Z(\lambda_0)$ &$\delta \mu^Z$& $\delta \mu_{GGD}$ & $\delta
\mu_{GDD}^{(1)}$& $(\delta
\mu_{GDD}^{(2)})$ & $\delta \mu_{\rm end}'$ & $\delta \mu_{\rm tot} $ &$\mu_{\rm th}$ &
$\tilde \mu_{\rm th}$\\

\noalign{\smallskip}\hline\noalign{\smallskip}

105 &5.675(5)&5.972&0.634 &-3.450 &2.980 & 0.155 &0.161  &-0.002  &-0.320  &5.652 &5.770\\

107 &5.585(8)&5.980&0.584 &-4.260 &3.680 & 0.186 &0.196  &-0.001  &-0.393  &5.587 &5.751\\

109 &5.538(4)&5.989&0.509 &-5.770 &4.990 & 0.247 &0.263  &-0.002  &-0.532  &5.457 &5.718\\

111 &5.503(7)&5.982&0.482 &-6.440 &5.550 & 0.286 & 0.229  &-0.002  &-0.605  &5.377 &5.691\\

113 &5.5289(2)&5.992&0.510&-5.770 &5.000 & 0.237 & 0.257  &-0.002  &-0.536  &5.456 &5.719\\

115 &5.5408(2)&6.004&0.609&-3.850 &3.330 & 0.155 & 0.168  &-0.001  &-0.364  &5.640 &5.782\\

117 &5.519(4)&6.016 &0.578&-4.390 &3.790 & 0.170 & 0.184  &-0.001  &-0.425  &5.590 &5.770\\

119 &5.515(1)&6.010 &0.614 &-3.780 &3.250 & 0.160 & 0.122  &-0.001  &-0.375  &5.635 &5.780\\

121 &5.502(5)&6.019 &0.634 &-3.470 &3.000 & 0.133 & 0.148  &-0.001  &-0.343  &5.676 &5.801\\

123 &5.491(7)&6.006 &0.674 &-2.900 &2.500 & 0.110 & 0.123  &-0.001  &-0.291  &5.715 &5.801\\

125 &5.502(9)&5.987 &0.734 &-2.160 &1.860 & 0.081 & 0.092  &   0.0  &-0.220  &5.767 &5.826\\

127 &5.522(8)&5.967 &0.780 &-1.500 &1.290 & 0.055 & 0.063  &   0.0  &-0.154  &5.813 &5.844\\

\noalign{\smallskip}\hline \hline

\end{tabular}\label{tab12}
\end{table*}

Let us now go to Sb isotopes. Here 4 experimental magnetic moments
in the ground states $5/2^+$ of the odd $^{115-121}$Sb isotopes are
known  which correspond to the $2d_{5/2}^p$ particle states and 7
moments for the $1g_{7/2}^p$ state in $^{121-133}$Sb isotopes, in
the $^{121}$Sb isotope we dealing with the first excited state. The
total magnetic moments are displayed in Fig. 24 whereas separate PC
corrections are represented in Table XIII. For the $1g_{7/2}^p$
state the TFFS without PC corrections describes the data better than
for In isotopes discussed above. The maximum of the disagreement,
$\simeq 0.3\mu_N$, is for $^{133}$Sb. Unfortunately, the PC
correction in the vicinity of the magic $^{132}$Sn nucleus, just as
for neighbors of $^{208}$Pb considered previously, is rather small.
This correction is rather small also for other isotopes, with
exception of two left members of the chain. Here the term $\delta
\mu'_{\rm end}$ is bigger than usual and leads to the noticeable
value of $\delta \mu_{\rm tot}$. It is worth to mention that the
$^{121}$Sb nucleus, $\lambda_0=1g_{7/2}$, is the only case with
positive value of $\delta \mu_{\rm tot}$. As usual, the ansatz
$\mu_{\rm th}\to \tilde \mu_{\rm th}$ damps the PC effects, in this
case resulting in better agreement with the data.

For $d_{5/2}$ states, the disagreement between TFFS and the data is
significant, about $0.6\;\mu_N$. The size of the PC correction
$\delta \mu_{\rm tot}$ is of this order or even bigger. Again the
use of the ansatz (\ref{mu-tild}) diminishes the PC effect and again
the experimental magnetic moments are between $\mu_{\rm th}$ and
$\tilde \mu_{\rm th}$ values.

Thus, we found that in semi-magic nuclei the PC corrections to the
magnetic moments are significant. If we exclude Bi isotopes where
they are small, the PC corrections have always the sign necessary to
improve the agreement with the data.

\begin{table*}[]
\caption{PC contributions  to the magnetic moments (in $\mu_N$) of
odd Sb isotopes. The notation is explained in the text nearby Table
IV.}

\begin{tabular}{c c c c c c c c c c c c c c}

\hline \hline\noalign{\smallskip} $A$ & $\lambda_0$  & $\mu_{\rm
exp}$ & $\mu_0$& $L^{\pi}$  &$Z(\lambda_0)$ & $\delta \mu^Z$&
$\delta \mu_{GGD}$ & $\delta \mu_{GDD}^{(1)}$& $(\delta
\mu_{GDD}^{(2)})$ & $\delta \mu_{\rm end}'$& $\delta \mu_{\rm tot} $ &$\mu_{\rm th}$ &$\tilde \mu_{\rm th}$\\

\noalign{\smallskip}\hline\noalign{\smallskip}

115 &$2d_{5/2}$  &3.46(1)&3.999 &$2^+$&0.484  &-4.260 &3.030 & 0.416 &-0.657  &0.004  &-0.819  &3.180 &3.602\\

117 &$2d_{5/2}$  &3.43(6)&3.993 &$2^+$&0.582  &-2.870 &2.030 &0.274 &-0.435  &0.002  &-0.565  &3.429 &3.665\\

119 &$2d_{5/2}$  &3.45(1)&4.008 &$2^+$&0.547  &-3.320 &2.350 &0.302 &-0.479  &0.003  &-0.668  &3.340  &3.643\\

121 &$2d_{5/2}$  &3.3654(3)&4.025 &$2^+$&0.591  &-2.780 &1.280 &0.252 &-0.003  &0.002  &-1.250  &2.775  &3.285\\

121 &$1g_{7/2}$  &2.518(7) &2.604 &$2^+$&0.607  &-1.690  &1.440 &0.232 &-0.224  &0.189  &0.174  &2.778  &2.710 \\

123 &$1g_{7/2}$  &2.5498(2)&2.619 &$2^+$&0.591  &-1.810  &1.390 &0.260 &-0.242  &-0.191 &-0.355 &2.264  & 2.409 \\

125 &$1g_{7/2}$  &2.63(4)  &2.633 &$2^+$&0.651  &-1.410  &1.080 &0.188 &-0.182  &-0.059 &-0.204 &2.430  &2.501  \\

127 &$1g_{7/2}$  &2.697(6) &2.652 &$2^+$&0.725  &-1.010  &0.773 &0.127 &-0.125  &-0.027 &-0.133 &2.519  &2.556 \\

129 &$1g_{7/2}$  &2.79(2)  &2.648 &$2^+$&0.798  &-0.670  &0.518 &0.081 & -0.079 &-0.012 &-0.084 &2.564  &2.581 \\

131 &$1g_{7/2}$  &2.89(1) &2.676 &$2^+$&0.836   & -0.523 &0.391 &0.089 & -0.057 &-0.003 &-0.046 &2.630  &2.638 \\

133 &$1g_{7/2}$  &3.00(1) &2.689 &$2^++3^-$&0.968 &-0.087 &0.060 &0.012  &-0.010 &0.006  &-0.009 &2.680  &2.680 \\

\noalign{\smallskip}\hline \hline

\end{tabular}\label{tab13}
\end{table*}

\begin{figure}[]
\vspace{10mm} \centerline {\includegraphics [width=86mm]{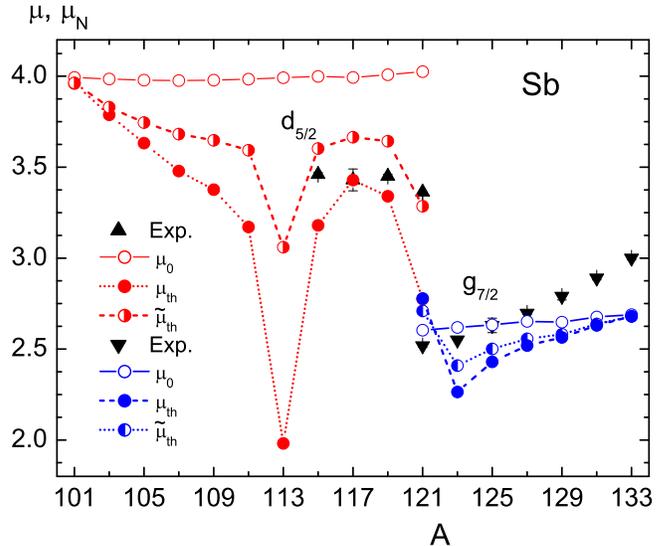}}
\vspace{2mm} \caption{Magnetic moments of Sb isotopes.}
\end{figure}

\section{Conclusion}

Within the self-consistent TFFS, we developed a model to calculate
the corrections to magnetic moments due to coupling to the low-lying
phonons in odd magic  and  semi-magic nuclei. The main idea of the
model is to consider explicitly only those PC diagrams which are
sensitive to the nucleus under consideration and the state
$\lambda_0$ of the odd nucleon. The omitted diagrams are supposed to
be included into the universal TFFS parameters.

Among semi-magic nuclei we limit ourselves to those where  the odd
nucleon belongs to the non-superfluid sub-system. This simplifies
the consideration significantly as all PC corrections can be
calculated without taking into account the pairing effects. The
perturbation theory up to the order of $g_L^2$ is developed, where
$g_L$ is the phonon-particle coupling vertex. The method guarantees
the total angular momentum conservation. For this aim, an ansatz is
proposed which takes into account the so-called tadpole term.

Three kinds of the PC corrections are considered. The end
correction, see Fig. 3, is the first one. The main (pole) part of
the end correction, $\delta \mu^Z$, results in $\sqrt{Z}$-factors on
the ``ends' of the effective field $V_{\lambda'\lambda}$. The second
one denoted as $\delta \mu_{GGD}$, see Fig. 5, describes the effect
of the induced interaction due to the $L$-phonon exchange. The third
one labeled as $\delta \mu_{GDD}$, see Fig. 6, describes the effect
of the magnetic moment of the phonon. This correction contains two
terms of different analytical structure. The term $\delta
\mu_{GDD}^{(1)}$ is regular at small phonon excitation energy
$\omega_L$ whereas the term $\delta \mu_{GDD}^{(2)}$ is singular at
$\omega_L \to 0$. Each of the three types of PC corrections has a
tadpole-like counterpart and the one shown in Fig. 7, corresponding
to the $\delta \mu_{GDD}$ term, is obviously the most important one.
In particular, it possesses the same singularity as the $\delta
\mu_{GDD}^{(2)}$ term. We suppose that the sum of tadpole-like terms
and the $\delta \mu_{GDD}^{(2)}$ one compensate each other. It is
motivated by the observation that such an ansatz leads to the total
angular momentum conservation with $g_L^2$ accuracy.

In magic nuclei, the main PC correction comes from the $3^-_1$
phonon, whereas in non-magic ones the $2^+$-phonon dominates. The
main observation of our calculations is the very strong cancelation
of the first two corrections, $\delta \mu^Z$ and $\delta \mu_{GGD}$,
The scale of the PC corrections may be characterized with the
quantity $(1-Z)$ which determines directly the term $\delta \mu^Z$.
With  only one exception, the total correction to the magnetic
moment $\delta \mu_{\rm tot}$ has the same sign as $\delta \mu^Z$,
and the absolute value is  up to ten times smaller than the
individual contributions.  As the result of this cancelation, in
magic nuclei where the value of $(1-Z)\simeq 0.1$ is typical, PC
corrections are as a rule negligible.  In any case, it is so for all
magnetic moments where experimental data exist. In particular, PC
corrections can not help to solve the old for the TFFS theory
problem of the magnetic moment of the ground state of $^{209}$Bi
where the theoretical value deviates by
  $\simeq 0.5\;\mu_N$ from the experiment. Evidently, the standard form
of the spin-dependent LM amplitude is not sufficient in this case
and additional terms including the spin-orbit and tensor in the
cross-channel interaction should be taken into account. Thus the
problem should be solved at the RPA level.

Simultaneously we calculated the gyromagnetic ratios $g_L^{\rm ph}$
of all low-lying phonons in $^{208}$Pb.  For the $3^-_1$ state it is
rather close to the BM model prediction $g_L^{\rm ph}=Z/A$ which is
a result of its strong collectivity. Other $L$-phonons are much less
collective and their gyromagnetic ratios strongly differ from the BM
ones.

We calculated the PC corrections to four chains proton-odd
semi-magic nuclei. The odd Tl and Bi isotopes are the odd neighbors
of the even lead isotopes. The Bi chain contains 7 nuclei with known
magnetic moment in the same $1h_{9/2}^p$ ground state and the PC
correction $\delta \mu_{\rm tot}$ is much less than  necessary to
explain the data in non-magic isotopes too. Evidently, the problem
of $^{209}$Bi should be solved first at the RPA level which will
reduce the disagreement for the whole Bi chain. For Tl isotopes
there are 10 isotopes with known magnetic moment in the
 $3s_{1/2}^{-p}$ ground state. For the magic isotope $^{207}$Tl TFFS
 agrees with the data perfectly well and the PC correction as it was
 said above is negligible. For non-magic Tl isotopes the typical
 disagreement is $\simeq 0.2\;\mu_N$ and PC corrections reduces
 it by one half. For odd-proton neighbors of the even tin
 isotopes, the odd In and Sb chains, the PC effects are stronger
 than in the lead chain. In such a situation, sometimes the use of
 the perturbation theory in $g_L^2$ seems questionable and we
 suggested the ansatz given in Eq. (\ref{mu-tild}) of an approximate
 account for the higher order  terms in $g_L^2$. It corresponds to partial
 summation of the pole end diagrams  in Fig. 3 itself and in the ``ends'' of Fig. 5 and Fig.
 6 as well. This ansatz, $\mu_{\rm th}\to \tilde \mu_{\rm th}$,
reduces the PC effect that is important in several cases. In
general, in both the chains  the PC corrections improves the
agreement.

\section{Acknowledgment}
The work was partly supported by the DFG and RFBR Grants Nos.
436RUS113/994/0-1 and 09-02-91352NNIO-a, by the Grant NSh-215.2012.2
of the Russian Ministry for Science and Education, and by the RFBR
Grants  11-02-00467-a,  12-02-00955-a, and 13-02-00085-a. Four of
us, E. S., O. A. , S. Ka., and S. T., are grateful to the Institut
f\"ur Kernphysik, Forschungszentrum J\"ulich for its hospitality.

{}


\begin{thebibliography}{}
\bibitem{Bel1} S. T. Belyev, Mat. Fys. Medd. Dan. Vid. Selsk. {\bf 31}, No. 11 (1959).

\bibitem{HFB-17} S. Goriely, N. Chamel, and J. M. Pearson, Phys. Rev.
Lett. {\bf 102}, 152503 (2009).

\bibitem{EKR} J. Erler, P. Kl\"{u}pfel, and P.-G. Reinhard, Phys. Rev. C {\bf 82},  044307 (2010).

\bibitem{Gogny} S. Goriely, S. Hilaire, M. Girod, and S. P\'eru, Phys. Rev.
Lett. {\bf 102}, 242501 (2009).

\bibitem{RHF}  P. Ring, Prog. Part. Nucl. Phys. {\bf 37}, 193 (1996).

\bibitem{Fay}
S. A. Fayans, S. V. Tolokonnikov, E. L. Trykov, and D. Zawischa, Nucl.
Phys. A {\bf 676}, 49 (2000).

\bibitem{KhS} V. A. Khodel and E. E. Saperstein, Phys. Rep. {\bf 92}, 183 (1982).

\bibitem{AB1} A. B. Migdal, {\it Theory of finite Fermi systems and applications to
atomic nuclei} (Nauka, Moscow, 1965; Wiley, New York, 1967).

\bibitem{Fay-Khod} S. A. Fayans and V. A. Khodel, JETP Lett.  {\bf 17},
444 (1973).

\bibitem{Bel2} S. T. Belyev,  A. V. Smirnov, S. V. Tolokonnikov, and S. A. Fayans, Yad. Fiz. {\bf 45},
1263 (1987) [Sov. J. Nucl. Phys. 45, 783 (1987)].

\bibitem{Bertsch1} J. Terasaki, J. Engel, and G. F. Bertsch, Phys. Rev. C {\bf 78}, 044311 (2008).

\bibitem{Vor} A. P. Severyukhin, V. V. Voronov, and Nguyen Van Giai, Phys. Rev. C {\bf 77}, 024322 (2008).

\bibitem{Bertsch2} G. F. Bertsch,  M. Girod, S. Hilaire, J.-P. Delaroche, H. Goutte,
and S. P\'eru, Phys. Rev. Lett. {\bf 99}, 032502 (2007).

\bibitem{RHF-RPA} A. Ansari and P. Ring, Phys. Rev. C {\bf 74}, 054313
(2006).

\bibitem{BE2} S. V. Tolokonnikov, S. Kamerdzhiev, D. Voytenkov, S. Krewald, and E. E. Saperstein,
Phys. Rev. C {\bf 84}, 064324 (2011).

\bibitem{solov} V. G. Soloviev, {\it Theory of Complex Nuclei}, (Moscow: Nauka, 1971;
Oxford: Pergamon Press, 1976).

\bibitem{NFT} P. F. Bortignon, R. A. Broglia, D. R. B\'es, and R. Liotta, Phys. Rep. {\bf 30}, 305, 1977.

\bibitem{colo1994} G. Colo, Nguyen Van Giai, P. F. Bortignon, R. A. Broglia, Phys. Rev.
C \textbf{50},1496 (1994).

\bibitem{revKST} S. Kamerdzhiev, J. Speth, G. Tertychny, Phys. Rep. \textbf{393},
1 (2004).

\bibitem{sarchi} D. Sarchi, P. F. Bortignon, G. Colo, Phys. Lett. \textbf{ B601}, 27
(2004).

\bibitem{tsel2007} V. Tselyaev, Phys. Rev. C \textbf{75}, 024306 (2007).

\bibitem{avd2007} A. Avdeenkov, F. Gruemmer, S. Kamerdzhiev {\it et al.}, Phys. Lett.
{\bf B653},196 (2007).

\bibitem{Pygmy} N. Paar, D. Vretenar, E. Khan, and Gianluca Colo,  Rep. Prog. Phys. {\bf 70}, 691 (2007).

\bibitem{PRL} N. Lyutorovich, V. Tselyaev, J. Speth, S. Krewald,  F. Gr\"ummer,
and P.-G. Reinhard,   Phys. Rev. Lett. {\bf 109}, 092502 (2012).

\bibitem{AGKK} A. Avdeenkov, S. Goriely, S. Kamerdzhiev, and S. Krewald, Phys. Rev. C {\bf 83}, 064316 (2011).

\bibitem{Stone} N. J. Stone,  Atomic Data Nuclear Data Table {\bf 90}, (2005) 75.

\bibitem{Q_Cu} P. Vingerhoets, K. T. Flanagan, and M. Avgoulea, {\it et. al.},
Phys. Rev. C {\bf 82}, 064311 (2010).

\bibitem{BABr}
M. Honma, T. Otsuka, B. A. Brown, and T. Mizusaki, Phys. Rev. C {\bf
69}, (2004) 034335.

\bibitem{mu1}  I.N. Borzov, E.E. Saperstein, and S.V. Tolokonnikov,
Phys. At. Nucl. {\bf 71}, 469 (2008).

\bibitem{mu2}  I. N. Borzov, E. E. Saperstein, S. V. Tolokonnikov, G. Neyens, and N. Severijns,
Eur. Phys. J. A {\bf 45}, 159 (2010).

\bibitem{QEPJ} S. V. Tolokonnikov, S. Kamerdzhiev, S. Krewald, E. E. Saperstein, and D. Voitenkov,
EPJA {\bf 48}, 70 (2012).

\bibitem{QEPJ-Web} S. Kamerdzhiev, S. Krewald, S. Tolokonnikov, E. E. Saperstein, and
D.Voitenkov.  EPJ Web of Conferences 38, 10002 (2012).

\bibitem{Q2pl} D. Voitenkov,  S. Kamerdzhiev, S. Krewald, E. E. Saperstein, and S. V. Tolokonnikov.
PRC {\bf 85}, 054319 (2012).

\bibitem{Fay1} A. V. Smirnov,  S. V. Tolokonnikov, and S. A. Fayans, Sov. J.  Nucl.
Phys. {\bf 48}, 995 (1988).

\bibitem{Fay5} S. A. Fayans, JETP Letters {\bf 68}, 169 (1998).

\bibitem{Tol-Sap} S. V. Tolokonnikov and E. E. Saperstein, Phys. Atom. Nucl. {\bf 73}, 1684 (2010).

\bibitem{BE2-Web}S. V. Tolokonnikov, S. Kamerdzhiev,  S. Krewald, E. E. Saperstein
and D. Voitenkov.  EPJ Web of Conferences 38, 04002 (2012).

\bibitem{Ba-Sp} R. Bauer, J. Speth, V. Klemt,P. Ring, E. Werner and T. Yamasaki, Nucl. Phys. A {\bf
209}, 535 (1973).

\bibitem{TBK} V. N. Tkachev, I.N. Borzov, S. P. Kamerdzhiev Sov. J. Nucl. Phys. {\bf 24},
373 (1976).

\bibitem{BTF} I. N. Borzov, S. V. Tolokonnikov, and S. A. Fayans, Yad. Fiz. {\bf 40},
1151 (1984) [Sov. J. Nucl. Phys. {\bf 40}, 732 (1984)].

\bibitem{Hamam} I.~Hamamoto, Phys. Lett. \textbf{B61}, (1973) 343.

\bibitem{Plat-mu} A. P. Platonov, Sov. J. Nucl. Phys. {\bf 34}, 342 (1981).

\bibitem{Tsel} V. N. Tselyaev, Sov. J. Nucl. Phys. {\bf 50}, 780
(1989).

\bibitem{Towner} I. S. Towner, Phys. Rep. {\bf 155}, 264 (1987).

\bibitem{PlSap} A. P. Platonov and E. E. Saperstein, Nucl. Phys. A {\bf 486}, (1988) 63.

\bibitem{ShB} S. Shlomo and G. F. Bertsch, Nucl. Phys. A {\bf 243}, (1975)
507.

\bibitem{scat_1} J. E. Wise, J. R. Calarco, J. P. Connely, S. A. Fayans,
F. W. Hersman, J. H. Heisenberg, R. S. Hicks, W. Kim, T. E.
Milliman, R. A. Miskimen, G. A. Peterson, A. P. Platonov, E. E.
Saperstein and R. P. Singhal,  Phys. Rev. C {\bf 47} (1993) 2539.

\bibitem{scat_2}
A. P. Platonov, E. E. Saperstein, S. V. Tolokonnikov, and S. A.
Fayans. Phys. At. Nuc. {\bf 58},  556  (1995).

\bibitem{BM1} A. Bohr and B. R. Mottelson, {\it Nuclear Structure} (Benjamin,
New York, Amsterdam, 1969.), Vol. 1.

\bibitem{Z-fac1}  E. E. Saperstein and S. V. Tolokonnikov, JETP Lett. {\bf 68},  553 (1998).

\bibitem{Z-fac2}  E. E. Saperstein and S. V. Tolokonnikov, Yad. Fiz. {\bf 62}, 1383 (1999)
[Phys. Atom. Nucl. {\bf 62}, 1302 (1999)].

\bibitem{BM2} A. Bohr and B. R. Mottelson, {\it Nuclear Structure} (Benjamin,
New York, Amsterdam, 1974.), Vol. 2.

\bibitem{Kaev2011}
S. P. Kamerdzhiev, A. V. Avdeenkov and D. A. Voitenkov, Yad. Fiz.
{\bf 74}, 1509 (2011) [Phys. Atom. Nucl. {\bf 74}, 1478 (2011)].


\bibitem{Kam-Sap} S. Kamerdzhiev and E. E. Saperstein,
Eur. Phys. J. {\bf A 37}, 333 (2008).

\bibitem{KaV2011} S. Kamerdzhiev and D. Voitenkov, arXiv:1110.0654 (2011).

\bibitem{Kaev83} S. P. Kamerdzhiev, Sov. J. Nucl. Phys. {\bf 38},188 (1983).


\bibitem{DmT} V. F. Dmitriev and V. B. Telicin, Nucl. Phys. A {\bf 402}, (1983) 581.


\end{thebibliography}
\end{document}